\documentclass[aps,prd,reprint,twoside,superscriptaddress,nofootinbib,eqsecnum,floatfix]{revtex4-2}

\usepackage{multirow}
\usepackage{array}
\usepackage{graphicx}
\usepackage{xcolor}
\definecolor{paperlinkblue}{RGB}{30,30,165}
\definecolor{paperlinkred}{RGB}{165,30,30}
\definecolor{paperfilemagenta}{RGB}{165,30,165}
\usepackage{hyperref}
\usepackage{pdfrender}
\hypersetup{
    colorlinks=true, 
    hypertexnames=false,
    linkcolor=paperlinkred,  
    citecolor=paperlinkblue,  
    filecolor=paperfilemagenta, 
    urlcolor=paperfilemagenta,   
    pdftitle={Constraining initial orbital eccentricity of inspiral-dominated gravitational-wave events with an analytic approximant},
    pdfauthor={Hemantakumar Phurailatpam, Achamveedu Gopakumar, Maria Haney, Tjonnie Li, and Srishti Tiwari},
    pdfsubject={doi:10.1103/94fx-wnjx url:https://doi.org/10.1103/94fx-wnjx},
    pdfkeywords={doi:10.1103/94fx-wnjx url:https://doi.org/10.1103/94fx-wnjx}
}
\usepackage{epsfig}
\usepackage{mathtools}
\usepackage{verbatim}
\usepackage{euscript}
\usepackage{amsthm,dsfont,amsfonts,amsmath,amssymb}
\usepackage{newtxtext,newtxmath}
\usepackage[cal=cm]{mathalpha}
\SetMathAlphabet{\mathcal}{bold}{U}{cmsy}{m}{n}

\usepackage{courier}
\usepackage{soul}
\usepackage{euscript,color,fontenc,textcomp,relsize}
\usepackage{bm,url,float,cleveref}
\usepackage{color,soul}
\usepackage[caption=false]{subfig}
\usepackage[utf8]{inputenc}
\usepackage[expansion=false]{microtype}
\NewCommandCopy{\prdtexttt}{\texttt}
\ExplSyntaxOn
\seq_new:N \l_prd_typewriter_words_seq

\DeclareRobustCommand{\texttt}[1]{%
  \prdtexttt{%
    \seq_set_split:Nnn \l_prd_typewriter_words_seq { ~ } {#1}
    \seq_map_indexed_inline:Nn \l_prd_typewriter_words_seq
      {
        \int_compare:nNnT {##1} > {1} {\space}
        \mbox{%
          \textpdfrender{
            TextRenderingMode=FillStroke,
            LineWidth=.2pt
          }{##2}%
        }
      }
  }%
}
\ExplSyntaxOff
\usepackage{orcidlink}
\usepackage{fancyhdr}
\usepackage{capt-of}
\definecolor{prdtext}{RGB}{35,31,32}
\definecolor{prdblue}{RGB}{46,48,146}
\hypersetup{
    linkcolor=paperlinkred,
    citecolor=paperlinkblue,
    filecolor=paperfilemagenta,
    urlcolor=paperfilemagenta
}
\makeatletter
\AtBeginDocument{%
  \ifdefined\pdfcolorstackinit
    \global\chardef\main@pdfcolorstack=\pdfcolorstackinit page direct
      {0.13725 0.12157 0.12549 rg 0.13725 0.12157 0.12549 RG}\relax
  \fi
  \color{prdtext}%
  \global\let\default@color\current@color
  \global\let\XC@default@color\XC@current@color
}
\makeatother

\allowdisplaybreaks
\newcommand{\prdwideopeningrule}{%
  \rule{244.97bp}{0.227bp}%
  \llap{\smash{\rule{0.227bp}{7.994bp}}}%
}
\newcommand{\prdwideclosingrule}{%
  \rlap{\smash{\rule[-7.767bp]{0.227bp}{7.994bp}}}%
  \rule{244.97bp}{0.227bp}%
}

\AtBeginDocument{%
  \thinmuskip=3mu
  \medmuskip=4mu
  \thickmuskip=5mu
}

\makeatletter
\DeclareRobustCommand{\prdsoftware}[1]{%
  {\normalfont\fontsize{8bp}{\baselineskip}\selectfont #1}%
}
\newlength{\prddisplayskip}
\newcommand{\prd@setdisplayskips}{%
  \abovedisplayskip=\prddisplayskip
  \belowdisplayskip=\prddisplayskip
  \abovedisplayshortskip=\prddisplayskip
  \belowdisplayshortskip=\prddisplayskip
}
\renewcommand\normalsize{%
  \@setfontsize\normalsize{10.5bp}{12.02bp}%
  \spaceskip .25em plus .16em minus .04em
  \prd@setdisplayskips
  \let\@listi\@listI
}
\renewcommand\small{\@setfontsize\small{9.5bp}{11bp}\spaceskip\z@skip\prd@setdisplayskips}
\renewcommand\footnotesize{\@setfontsize\footnotesize{9.5bp}{11bp}\spaceskip\z@skip\prd@setdisplayskips}
\normalsize
\def\frontmatter@abstractwidth{400\p@}
\def\frontmatter@abstractfont{%
  \fontsize{9.5bp}{12bp}\selectfont
  \spaceskip\z@skip
  \parindent1em\relax
  \adjust@abstractwidth
}
\def\frontmatter@title@above{\vspace*{13.8\p@}}
\def\frontmatter@title@format{\fontsize{12.5bp}{14bp}\selectfont\spaceskip\z@skip\bfseries\centering\parskip\z@skip}
\def\frontmatter@preabstractspace{8.5\p@}
\def\frontmatter@postabstractspace{30\p@}
\def\frontmatter@RRAPformat#1{%
  \removelastskip\vspace{6bp}%
  {\small\centering
  (Received 19 August 2025; accepted 8 July 2026; published 3 September 2026)\par}%
}
\def\section{\@startsection{section}{1}{\z@}{15bp plus4bp minus4bp}{6bp}%
  {\normalfont\normalsize\bfseries\boldmath\centering}}
\def\@seccntformat#1{\csname the#1\endcsname.\hskip0.3333em\relax}
\def\subsection{\@startsection{subsection}{2}{\z@}{13bp plus4bp minus4bp}{5bp}%
  {\normalfont\normalsize\bfseries\boldmath\centering}}
\def\subsubsection{\@startsection{subsubsection}{3}{\z@}{13bp plus4bp minus4bp}{5bp}%
  {\normalfont\normalsize\itshape\centering}}
\skip\footins=20bp plus2bp minus2bp
\def\footnoterule{\kern-4bp\hrule width36bp height0.51022bp\kern4bp}
\long\def\prd@makefntext#1{%
  \def\baselinestretch{1}%
  \leftskip\z@\rightskip\z@\parindent10.5bp
  \noindent\nobreak\hskip10.5bp\@makefnmark#1\par
}
\let\@makefntext\prd@makefntext
\long\def\frontmatter@makefntext#1{%
  \prd@makefntext{%
    \Hy@raisedlink{\hyper@anchorstart{frontmatter.\expandafter\the\csname c@\@mpfn\endcsname}\hyper@anchorend}%
    #1}%
}
\def\@hangfrom@appendix#1#2#3{%
  #1\MakeTextUppercase{#2\@if@empty{#3}{}{:\ #3}}%
}
\AtBeginDocument{\let\NAT@space\@empty}
\AtBeginDocument{%
  \def\bibfont{\fontsize{9.5bp}{11.25bp}\selectfont\spaceskip\z@skip
    \@clubpenalty\clubpenalty}%
  \setlength{\bibsep}{-0.25bp}%
  \def\bib@device#1#2{\hb@xt@#1{\hfil\rule{245.99bp}{0.28348bp}\hfil}}%
  \patchcmd{\bibsection}{\addvspace{19\p@}}{\addvspace{30\p@}}{}%
    {\PackageError{prd-layout}{Bibliography spacing patch failed}{Check REVTeX's bibsection definition.}}%
}
\def\@caption@fignum@sep{.\nobreak\hskip9.5bp\relax}
\makeatother

\fancypagestyle{prdarticle}{%
  \fancyhf{}%
  \fancyhead[LE]{\normalsize HEMANTAKUMAR PHURAILATPAM \textit{et al.}}%
  \fancyhead[LO]{\normalsize\MakeUppercase{Constraining initial orbital eccentricity of \ldots}}%
  \fancyhead[R]{\normalsize PHYS. REV. D \textbf{114}, 063003 (2026)}%
  \fancyfoot[C]{\fontsize{9bp}{11bp}\selectfont 063003-\thepage}%
}
\fancypagestyle{prdfirstpage}{%
  \fancyhf{}%
  \fancyhead[C]{\fontsize{12bp}{14bp}\selectfont PHYSICAL REVIEW D \textbf{114}, 063003 (2026)}%
  \fancyfoot[L]{\fontsize{9bp}{11bp}\selectfont 2470-0010/2026/114(6)/063003(31)}%
  \fancyfoot[C]{\fontsize{9bp}{11bp}\selectfont 063003-1}%
  \fancyfoot[R]{\fontsize{9bp}{11bp}\selectfont \textcopyright\ 2026 American Physical Society}%
}
\begin{document}
\color{prdtext}
\raggedbottom

\title{Constraining initial orbital eccentricity of inspiral-dominated gravitational-\texorpdfstring{\\}{ }wave events with an analytic approximant}

\author{Hemantakumar Phurailatpam\,\orcidlink{0000-0002-0471-3724}}
\email[Contact author: ]{hemantaph@link.cuhk.edu.hk}
\affiliation{Department of Physics, \href{https://ror.org/00t33hh48}{The Chinese University of Hong Kong}, Shatin, New Territories, Hong Kong}

\author{Achamveedu Gopakumar\,\orcidlink{0000-0003-4274-4369}}
\affiliation{Department of Astronomy and Astrophysics, \href{https://ror.org/03ht1xw27}{Tata Institute of Fundamental Research}, Mumbai 400005, India}

\author{Maria Haney\,\orcidlink{0000-0001-7554-3665}}
\affiliation{\mbox{\href{https://ror.org/00f9tz983}{Nikhef---National Institute for Subatomic Physics}, Science Park 105, 1098 XG Amsterdam, The Netherlands}}

\author{Tjonnie Li\,\orcidlink{0000-0003-4297-7365}}
\affiliation{Leuven Gravity Institute, \href{https://ror.org/05f950310}{KU Leuven}, Celestijnenlaan 200D box 2415, 3001 Leuven, Belgium}
\affiliation{Department of Physics and Astronomy, Laboratory for Semiconductor Physics, \href{https://ror.org/05f950310}{KU Leuven},\\ B-3001 Leuven, Belgium}
\affiliation{\mbox{\href{https://ror.org/05f950310}{KU Leuven}, Department of Electrical Engineering (ESAT), STADIUS Center for Dynamical Systems, Signal Processing}\\ and Data Analytics, B-3001 Leuven, Belgium}

\author{Srishti Tiwari}
\affiliation{\href{https://ror.org/04etcj997}{The Inter-University Centre for Astronomy and Astrophysics}, Post Bag 4, Ganeshkhind, Pune 411007, India}

\received{19 August 2025}
\accepted{8 July 2026}
\published{3 September 2026}
\begin{abstract}
The LIGO-Virgo-KAGRA consortium has sporadically detected inspiral-dominated gravitational-wave events such as GW170817 and GW190425. These events offer an opportunity to constrain possible initial (residual) orbital eccentricities using purely inspiral template families. We detail the implementation of an \prdsoftware{LALSuite} approximant, \texttt{TaylorF2Ecck}, which analytically models inspiral gravitational waves from nonspinning compact binaries in post-Newtonian (PN)-accurate eccentric orbits and restricts the initial-eccentricity contributions to leading order. Specifically, our frequency-domain approximant consistently incorporates orbital, advance of periastron, and gravitational-wave emission effects fully up to 3PN order. We conduct detailed parameter-estimation studies of GW170817 and GW190425 using \texttt{TaylorF2Ecck}, following comprehensive sanity checks to validate model performance and investigate the influence of eccentricity and periastron advance in the relevant parameter space. The results indicate that the initial eccentricity at 20~Hz is negligible within the 90\% credible intervals (less than 0.016 for GW170817 and less than 0.023 for GW190425, with both posterior distributions railing toward zero), and Bayes factors show no strong evidence favoring the eccentric waveform over the quasicircular waveform. At such negligible initial eccentricities, comparisons between eccentric models with and without periastron advance show no clear signature of this effect, with no significant model-dependent shifts in the posterior distributions and no strong Bayes-factor evidence favoring one model over the other. Additionally, these detailed studies reveal the importance of incorporating initial-eccentricity contributions at least up to 3.5PN order and discuss its implications. We substantiate this inference by employing versions of the quasicircular \texttt{TaylorF2} approximant that incorporate Fourier phase contributions beyond the conventional 3.5PN order.

\vspace{6pt}
\noindent DOI: \href{https://doi.org/10.1103/94fx-wnjx}{10.1103/94fx-wnjx}
\end{abstract}

\maketitle
\thispagestyle{prdfirstpage}
\normalsize


\section{Introduction}\label{sec:level1}

The LIGO-Virgo-KAGRA (LVK) Collaboration routinely detects transient gravitational-wave (GW) events from compact binary coalescences, predominantly those of stellar-mass black hole binaries (BBHs)~\cite{Abbott_2020, AasiJ2015, Acernese_2014, ptaa125, PhysRevD.88.043007, GraceDBO4, GWTC5, GWTC5_pop}. With the GWTC-5 catalog now containing 390 candidates following the O4b release~\cite{GWTC5, GWTC5_pop}, the era of GW astronomy is firmly established~\cite{LIGOScientific:2018mvr, Abbott:2020niy, LIGOScientific:2021usb, LVK_O3c}. Additionally, the consortium has sporadically detected GWs from a few neutron star-neutron star (BNS) and neutron star-black hole (NSBH) binaries~\cite{LVK_GW170817, BHNSs, GWTC4, GWTC5} with long inspirals. These detections currently constrain local merger-rate densities to $\sim 7.6\text{--}250~\mathrm{Gpc}^{-3}\,\mathrm{yr}^{-1}$ for BNS systems and $\sim 9.1\text{--}84~\mathrm{Gpc}^{-3}\,\mathrm{yr}^{-1}$ for NSBH systems~\cite{O4pop, GWTC5_pop}. Furthermore, the discovery of inspiral GWs from the merging BNS binary GW170817, along with its electromagnetic counterparts, marked the advent of multimessenger astronomy~\cite{MM170817}. It is important to note that the currently operational GW observatories are primarily sensitive to the inspiral phase of GWs from BNS systems~\cite{LVK_GW170817}.

Such inspiral-dominated GW events are typically detected using the fully analytic, frequency-domain \texttt{TaylorF2} approximant~\cite{BIOPS} within the LVK Algorithm Library Suite (\prdsoftware{LALSuite})~\cite{lalsuite}. 
This approximant employs the post-Newtonian (PN) approximation to general relativity to model the temporally evolving GW polarization states associated with the inspiral phase of compact binary coalescence~\cite{LB_LR}.
Note that the PN approximation requires slow motion and weak fields, and hence requires $(v/c)^2 \ll 1 $ and $Gm/(r c^2) \ll 1$, where $v$, $m$, and $r$ are the orbital velocity, total mass, and relative separation of the binary, respectively.
The computationally efficient \texttt{TaylorF2} approximant models inspiral GWs from compact binaries in quasicircular orbits.
 The standard \prdsoftware{LALSuite} implementation typically employs Newtonian (quadrupolar) order amplitude corrections, while the Fourier phase corrections are accurate to 3.5PN order, providing $(v/c)^{7}$ order corrections to the dominant quadrupolar contributions~\cite{BIOPS}.
 This approximant utilizes the stationary phase approximation (SPA) to obtain analytical frequency-domain inspiral templates from their time-domain counterparts~\cite{DIS01}.
 It is possible to extend the approach to incorporate the effects of spin and orbital eccentricity~\cite{PW95,THG, Moore2019, Boetzel2019, CTGL21, Paul2023, Henry2023, sridhar2024}. 

Inspiral-dominated GW events, such as GW170817, spend substantially more time in the LVK frequency window (around 20~Hz to kHz) than typical BBH events. For instance, the inspiral phase of GW170817
spans approximately 168\footnote{Inspiral duration estimated using the \texttt{findchirp\_chirptime} function in \texttt{gwsnr}~\cite{phurailatpam2025}, from 20~Hz to the last stable orbit frequency $f_{\rm LSO}$ [see $f_{\rm LSO}$ expression in Eq.~(\ref{eq:flso}) of Appendix~\ref{sec:TaylorF2Ecck_hf}]. This is similar to the \texttt{calculate\_time\_to\_merger} function in \prdsoftware{Bilby} and \texttt{SimInspiralTaylorF2ReducedSpinChirpTime} in \texttt{LALSimulation}.\label{xx}}~s, while that of the first GW event, GW150914, lasted a mere 0.7~s (see Footnote~\ref{xx}). 
Consequently, inspiral-dominated LVK events are particularly well suited for exploring the effects of any residual orbital eccentricity, because they undergo a large number of GW cycles within the operational frequency windows of GW observatories~\cite{Favata2022}.
In general, typical LVK binaries are expected to exhibit small residual eccentricities, less than $10^{-4}$, although BNS systems may display higher initial eccentricities~\cite{Kowalska11}.
The small initial eccentricities are expected due to the efficiency of GWs in circularizing isolated eccentric binaries by rapidly radiating away angular momentum~\cite{Peters64}.
Astrophysical considerations suggest that the detection of residual eccentricities in LVK events can help distinguish between the two possible astrophysical formation channels for stellar-mass compact binaries~\cite{Zevin2021}.
In the isolated formation channel, compact binaries originate from the evolution of massive binary stars without external perturbation~\cite{Broekgaarden2021}, and astrophysical assessments indicate that their orbits should be circularized by the time their GW signals enter the LVK frequency window~\cite{Belczynski16}.
In contrast, compact binaries can evolve dynamically when BHs or NSs form binaries with other BHs or NSs in dense stellar environments~\cite{Sigurdsson1993}.
Such systems include nuclear star clusters, globular clusters, young star clusters, or even the discs of active galactic nuclei (for a detailed review, see Refs.~\cite{Sigurdsson1993,Miller2002,Rodriguez2016b,Antonini2016a, Antonini2016b,Askar2017, Haiman2009, Bartos2017}).
It is argued that the presence of measurable initial orbital eccentricity should aid in distinguishing between these scenarios~\cite{Antonini2014, Rodriguez2016a, Wen2003, Antognini2014,Gultekin2004, OLeary2006, Haster2016, DOrazio2018, Fragione2019, Zevin2021, Breivik2016}.
This understanding was the primary motivation behind early attempts to constrain the initial eccentricity of GW150914, as discussed in Ref.~\cite{GW150914_prop}. Significant work has gone into creating gravitational waveform models that incorporate the effects of orbital eccentricity~\cite{Morras2025a, planas2025b, RomeroShaw2019, Nitz2020,Romero_Shaw_2020_GW190521, gayathri2022, Wu2020, Favata2022, OShea2023, RomeroShaw2021, Gamba2022, Bonino2023, iglesias2023, Klein2010, Klein_2018, Klein2021}. Such models have been applied to analyze GW signals observed by LIGO and Virgo, and have also been used in recent studies to set bounds on initial eccentricities of a few LVK coalescence events~\cite{BBG23, Morras2025b, planas2025a}.

Firm bounds on the initial eccentricity $e_0$ for two inspiral-dominated LVK events, namely GW170817 and GW190425, were established in Ref.~\cite{LNB}.
This analysis utilized the \texttt{TaylorF2Ecc} approximant, which incorporates leading-order initial eccentricity contributions that are 3PN-accurate in the Fourier phase of the widely used quasicircular \texttt{TaylorF2} inspiral approximant~\cite{MF16}.
The \texttt{TaylorF2Ecc} approximant, as detailed in Refs.~\cite{Favata2022,MF16}, also includes 3.5PN-order circular inspiral contributions to the dominant Fourier phase component and features quadrupolar (circular) amplitudes. However, it does not account for periastron-advance effects, which enter the compact binary dynamics at 1PN order~\cite{DD85}. 
This effect refers to the gradual rotation of the closest point (periastron) of an eccentric orbit involving compact objects due to general relativistic effects.

Our current work introduces a new \prdsoftware{LALSuite}-compatible approximant, \texttt{TaylorF2Ecck}, which utilizes and adapts inputs from Ref.~\cite{TGHH}. Consequently, our effort provides fully analytic frequency-domain inspiral templates for eccentric inspirals, incorporating ${\cal O}(e_0^2)$ contributions arising from 3PN-accurate periastron advance and GW emission in the relevant Fourier phases, while their amplitudes are at the quadrupolar order and accurate to ${\cal O}(e_0)$.
In other words, our inspiral approximant extends the \texttt{TaylorF2} inspiral approximant for nonspinning compact binaries by consistently including all leading-order $e_0$ contributions that arise from orbital, periastron advance, and GW emission effects, with these contributions included in the phase up to 3PN order and in quadrupolar-order amplitude.
Therefore, our \texttt{TaylorF2Ecck} approximant provides a computationally efficient and accurate template family for analyzing nonspinning, low-eccentricity inspiral events.
We note in passing that this is a restricted version of the approximant presented in Ref.~\cite{TGHH}, which provides eccentric corrections accurate to ${\cal O}(e_0^6)$ with amplitude corrections at 1PN order.

In this paper, we employ our \texttt{TaylorF2Ecck} approximant to constrain $e_0$ for two inspiral-dominated GW events, GW170817 and GW190425.
Using Bayesian inference with nested sampling~\cite{Skilling2004}, we estimate $e_0$ for these events and find results consistent with negligible eccentricity at a gravitational-wave frequency of $20$~Hz, with the posterior distributions railing toward zero.
Bayes-factor comparisons with the quasicircular \texttt{TaylorF2} model show no strong evidence favoring the eccentric model over the quasicircular model.
Our analysis shows that these constraints remain robust under both uniform and log-uniform $e_0$ priors.
Furthermore, Bayes-factor evaluations show no strong preference for \texttt{TaylorF2Ecck}, which includes periastron advance, over \texttt{TaylorF2Ecc}, which does not.
At such negligible eccentricities, we also observe no significant model-dependent shifts in the posterior distributions due to periastron advance.
However, we identify a substantial bias in the mass-ratio posteriors when comparing these models against versions that include 3.5PN-order circular contributions to the Fourier phase.
Considering the known correlation between mass and eccentricity, these detailed studies indicate the importance of incorporating initial eccentricity contributions at least up to 3.5PN order.
We substantiate this inference by employing variants of the quasicircular \texttt{TaylorF2} approximant that incorporate Fourier phase contributions beyond the conventional 3.5PN order.

This manuscript is structured as follows. Section~\ref{sec:Model} provides an overview of the frequency-domain waveforms employed in this work. In this section, we introduce the \texttt{TaylorF2Ecck} approximant and present waveform sanity checks. Section~\ref{sec:data_analysis} describes the data-analysis methods and presents parameter-estimation results for GW170817 and GW190425. Our constraints on $e_0$ for these events, together with their implications and limitations, are discussed in this section. Section~\ref{sec:conclusion} summarizes our main findings and discusses future extensions of this work. Appendixes~\ref{sec:TaylorF2Ecck_full} and~\ref{sec:TaylorF2Ecch_full} provide the explicit PN-accurate expressions for the inspiral approximants. Appendix~\ref{sec:sanity_tests_appendix} presents supplementary sanity checks. Appendix~\ref{sec:appendix_injection_recovery} provides supplementary injection-recovery results. Appendixes~\ref{sec:full_gw170817} and~\ref{sec:full_GW190425} show the full parameter-estimation corner plots for GW170817 and GW190425. Appendix~\ref{sec:comparison_pyefpe_taylorf2_family} compares \texttt{TaylorF2Ecck} and \texttt{TaylorF2Ecc} with the \texttt{pyEFPE} waveform model.

\section{Inspiral waveform approximants}\label{sec:Model}

In this section, we briefly introduce various fully analytic inspiral approximants employed in our investigations. These include existing approximants available in the \texttt{LALSimulation} library of \prdsoftware{LALSuite}, such as \texttt{TaylorF2} and \texttt{TaylorF2Ecc}, which are detailed in Refs.~\cite{BIOPS, MF16}, and two new eccentric inspiral approximants, namely \texttt{TaylorF2Ecck} and \texttt{TaylorF2Ecch}, influenced by Refs.~\cite{TGHH, THG, Yunes2009}. We first describe these approximants and then present detailed sanity checks for them.

\subsection{Eccentric inspiral approximants}\label{sec:TaylorF2Ecck_intro}
 
We begin by presenting our \texttt{TaylorF2Ecck} approximant, which employs expressions from Ref.~\cite{TGHH}. This fully analytic frequency-domain approximant models inspiral GWs from nonspinning compact point-mass binaries in precessing PN-accurate eccentric orbits. 
While Ref.~\cite{TGHH} offers expressions with 1PN-accurate amplitude corrections and 3PN-accurate phase evolution, both incorporating eccentricity effects up to ${\cal O}(e_0^6)$ at each PN order, our implementation for this work adopts a restricted version, primarily to reduce computational costs in parameter estimation. Specifically, we restrict our attention to a version that incorporates quadrupolar-order amplitudes, while the relevant Fourier phases are fully 3PN-accurate to leading order in ${\cal O}(e_0^2)$ contributions, and remain ${\cal O}(e_0)$ in the Fourier amplitude.
Our \texttt{TaylorF2Ecck} approximant, adapted from Eq.~(2.5) of Ref.~\cite{TGHH}, reads
\begin{subequations}
\begin{align}
\tilde{h}(f) = & \, {\cal \tilde{A}} \left(\frac{Gm\pi f}{c^3}\right)^{-7/6} \notag \\* 
& \times \sum_{j,n} \xi_{j,n} \left(\frac{j}{2}\right)^{2/3} e^{-i(\Psi_{j,n}-\pi/4)} \; \Theta_{j,n}\,,
\end{align}
\noindent with
\begin{align}
&{\cal \tilde{A}} =   \, -\left( \frac{5\pi \eta}{384} \right)^{1/2} \frac{(G m)^2}{c^5 d_L}\,,
\end{align}
\end{subequations}
\noindent where the $(j, n)$ harmonic combinations take values $(1,0)$, $(1,-2)$, $(2,-2)$, and $(3,-2)$ for the current version of the \texttt{TaylorF2Ecck} approximant. The $(2,-2)$ harmonic is the dominant one, while the remaining harmonics are subdominant, with their amplitudes suppressed at low $e_0$ values (see Eq.~(\ref{eq:xiEcck_simple})).
Following Ref.~\cite{TGHH}, ${\cal \tilde{A}}$ is the harmonic-independent amplitude prefactor and $\xi_{j,n}$ are the harmonic-dependent amplitude coefficients of our frequency-domain approximant.
Clearly, ${\cal \tilde{A}}$ depends on the luminosity distance $d_L$, the total mass ($m=m_1+m_2$), and the symmetric mass ratio ($\eta=\frac{m_1\,m_2}{m^2} = \frac{q}{(1+q)^2}$), where $m_1$ and $m_2$ are the individual masses and $q$ is the mass ratio. Note that it is possible to express the $m$- and $\eta$-dependence of ${\cal \tilde{A}}$ in terms of the more familiar chirp mass ${\cal M}$~\cite{findchirp}. In the above equation, $\Theta_{j,n}$ is the unit step function, which allows us to appropriately terminate each harmonic at the correct physically allowed frequency limit (see Eq.~(3.26) of Ref.~\cite{TGHH}).
Furthermore, $\xi_{j,n}$ depends on $e_0$, the orbital inclination $\iota$, the azimuthal angle $\beta$, and the antenna pattern functions $F_+$, $F_{\times}$. We set $\beta=0$, following \texttt{LALSimulation} conventions. In what follows, we list the expressions for $\xi_{j,n}$ accurate to $\mathcal{O}(e_0)$, influenced by Eq.~(3.16) of Ref.~\cite{TGHH}. They read,
\begin{subequations}\label{eq:xiEcck_simple}
\begin{align}
&\xi_{1,0} =\, F_+\left\{-\frac{\sin^2 \iota}{\chi^{19/18}} e_0\right\}\,, \\[9.5bp]
&\xi_{1,-2} =\, F_+\left\{-\frac{3(3+\cos(2\iota))}{4\chi^{19/18}} e_0 \right\} + i F_\times \left\{-\frac{\cos\iota}{\chi^{19/18}} e_0 \right\}\,,
\end{align}

\onecolumngrid
\begin{align}
&\xi_{2,-2} =\, F_+\left\{3 + \cos(2\iota)\right\} + i F_\times \left\{4 \cos\iota\right\}\,, \\[5.0bp]
&\xi_{3,-2} =\, F_+\left\{\frac{9(3+\cos(2\iota))}{4\chi^{19/18}} e_0 \right\} + i F_\times \left\{\frac{9 \cos\iota}{\chi^{19/18}} e_0 \right\}\,,
\end{align}
\end{subequations}
\noindent where the dimensionless parameter $\chi$ is given by $\chi = f/f_0$ (with $f_0$ being the initial GW frequency). Note that the dominant $(2,-2)$ mode lacks an $e_0$ correction and is circular in nature. 

Although our approximant is 3PN-accurate, for visualization purposes we now list the fully analytic Fourier phases for the $(j, n)$ harmonics restricted to 1PN order, including all $\mathcal{O}(e_0^2)$ corrections. This form for the $(j, n)$-th harmonic reads
\begin{subequations}
\label{eq:psi_ecck}
\begin{align}
\Psi_{j,n}= & \, -2 \pi  f t_c+ \phi_c \left(j- (j+n)\frac{k}{1+k}\right)-\frac{3 j}{256 \eta  x_{j,n}^{5/2}}\Biggl[ 1-\frac{2355 e_0^2}{1462 \chi ^{19/9}}+x_{j,n} \Biggl\{ \frac{55 \eta }{9}-\frac{25 n}{3 j}-\frac{2585}{756} \notag \\* & + e_0^2 \left( \frac{\frac{154645 \eta }{17544}-\frac{2223905}{491232}}{\chi ^{25/9}}+\frac{-\frac{128365 \eta }{12432}+\frac{1805 n}{172 j}+\frac{69114725}{14968128}}{\chi ^{19/9}} \right) \Biggl\} + \ldots\Biggr]\,, \label{eq:psi_simple}
\end{align}\label{eq:x_jn}
\par
\nointerlineskip
\hbox to\textwidth{\hfil\vrule height0pt depth6pt width0pt\prdwideclosingrule}
\vspace{4.0pt}\twocolumngrid
\noindent with
\begin{align}
&x_{j,n} =\, \left[\frac{2 \pi G m f}{\left\{j - (j+n)\frac{k}{1+k}\right\}c^3}\right]^{2/3}\,, 
\end{align}
\end{subequations}\\
\noindent and we refer to Eq.~(\ref{eq:Ecck_psi_full}) of Appendix~\ref{sec:TaylorF2Ecck_full} for the 3PN-accurate version.
In the above equation, $t_c$ and $\phi_c$ are the time and phase of coalescence, respectively, while $k$ is the rate of periastron advance and $x_{j,n}$ is the harmonic-dependent dimensionless PN parameter.
We emphasize that our {\it restricted} \texttt{TaylorF2Ecck} approximant contains {\it four} harmonics, whose Fourier phases are fully 3PN-accurate and incorporate $\mathcal{O}(e_0^2)$ corrections. We employ the 3PN-accurate version of $k$ in our approximant, given in Eq.~(\ref{eq:k_full}) of Appendix~\ref{sec:TaylorF2Ecck_full}. 

To isolate the effects of $k$, we now introduce another eccentric approximant, namely \texttt{TaylorF2Ecch}, influenced by Refs.~\cite{THG,Yunes2009}. This approximant omits the effects of periastron advance in both Fourier amplitudes and phases, although its Fourier phases remain 3PN-accurate and incorporate $\mathcal{O}(e_0^2)$ corrections. The omission of periastron advance from the amplitudes ensures that \texttt{TaylorF2Ecch} is expressed as a sum over single harmonics $j$, whereas \texttt{TaylorF2Ecck} includes mode splitting (as in the splitting of $j=1$ into $(j,n)=(1,0),(1,-2)$, as evident from Eq.~(\ref{eq:xiEcck_simple})). 
In other words, this approximant has reduced harmonic content, with modes limited to $j=1,2,3$, and lacks the additional phase shift associated with periastron effects found in the subdominant harmonics of \texttt{TaylorF2Ecck} (see the second term in Eq.~(\ref{eq:psi_simple}) for $\Psi_{j,n}$).
Therefore, this approximant should help delineate scenarios where periastron effects are significant relative to our \texttt{TaylorF2Ecck} templates, as discussed in Sec.~\ref{sec:sanity_check}. 
Our \texttt{TaylorF2Ecch} approximant, influenced by Ref.~\cite{THG}, reads

\begin{align}
\tilde{h}(f) = & \, {\cal \tilde{A}} \left(\frac{Gm\pi f}{c^3}\right)^{-7/6}\notag \\*
& \times \sum_{j} \xi_{j} \left(\frac{j}{2}\right)^{2/3} e^{-i(\Psi_{j}-\pi/4)}\; \Theta_{j}\,,
\end{align}

\noindent while the harmonic-dependent amplitude coefficients $\xi_{j}$ have the following structure.
\begin{subequations}\label{eq:xiEcch_simple}
\begin{align}
&\xi_{1} =\, F_+\left\{-\frac{3(3+\cos(2\iota))}{4\chi^{19/18}}e_0 - \frac{\sin(\iota)^2}{\chi^{19/18}}e_0\right\} \notag\\*
&\qquad+ i F_\times\left\{-\frac{\cos(\iota)}{\chi^{19/18}}e_0 \right\}\,, \\[15.0bp] 
&\xi_{2} =\, F_+\left\{3 + \cos(2\iota)\right\} + i F_\times\left\{4\,\cos(\iota)\right\}\,, \\[15.0bp] 
&\xi_{3} =\, F_+\left\{ \frac{9(3+\cos(2\iota))}{4\chi^{19/18}}e_0 \right\} + i F_\times \left\{\frac{9\,\cos(\iota)}{\chi^{19/18}}e_0 \right\}\,.
\end{align}
\end{subequations}
The 3PN-accurate expression of $\Psi_{j}$ is included in Appendix~\ref{sec:TaylorF2Ecch_full} (see Eq.~(\ref{eq:Ecch_psi_full})).
For comparison with Eq.~(\ref{eq:psi_simple}), the 1PN-accurate Fourier phase of the $j$-th harmonic is
\begin{align}
\Psi_{j}= & \, -2 \pi  f t_c+ \phi_c j -\frac{3 j}{256 \eta  x_j^{5/2}}\Biggl[ 1-\frac{2355 e_0^2}{1462 \chi ^{19/9}} \notag \\*
&+x_j \Biggl\{\frac{55 \eta }{9}+\frac{3715}{756} + e_0^2 \biggl(\frac{\frac{154645 \eta }{17544}-\frac{2223905}{491232}}{\chi ^{25/9}}\notag \\*
&+\frac{-\frac{128365 \eta }{12432}-\frac{2045665}{348096}}{\chi ^{19/9}}\biggr) \Biggr\} + \ldots\Biggr] \,. \label{eq:psi1PNEcch}
\end{align}

\begin{table}[t!]
\centering
\setlength{\abovecaptionskip}{0pt}
\caption{This table compares the inspiral approximants used in the current study. All approximants assume nonspinning compact binaries with quadrupolar-order amplitudes.
\texttt{TaylorF2Ecck} features 3PN-$\mathcal{O}(e_0^2)$ accurate Fourier phases, multiple harmonics beyond the (2,-2) mode, amplitude with $\mathcal{O}(e_0)$ corrections, and all the effects of periastron advance.
\texttt{TaylorF2Ecch} is similar to \texttt{TaylorF2Ecck} in Fourier phase and amplitude accuracy (PN and $e_0$ order) but omits all periastron-advance related effects. \texttt{TaylorF2Ecc} includes 3PN-$\mathcal{O}(e_0^2)$ accurate Fourier phase corrections but is restricted to the dominant harmonic with quadrupolar circular amplitude, while neglecting periastron effects~\cite{Favata2022}. Our quasicircular \texttt{TaylorF2} template family allows Fourier phases accurate to different PN orders.
}
\small
\renewcommand{\arraystretch}{1.0}
\setlength{\tabcolsep}{0pt}
\begin{tabular}{@{}p{70bp}>{\centering\arraybackslash}p{55bp}>{\centering\arraybackslash}p{49bp}>{\raggedright\arraybackslash}p{71bp}@{}}
\hline\hline
\mbox{}\newline Model & Fourier phase \newline accuracy & Amplitude \newline correction & \centering Harmonics \newline ($j, n$)\arraybackslash \\
\hline
\texttt{TaylorF2Ecck} & 3PN-$\mathcal{O}(e_0^2)$ & $\mathcal{O}(e_0)$ & $(1,0)$, $(1,-2)$,\newline $(2,-2)$, $(3,-2)$ \\
\texttt{TaylorF2Ecch} & 3PN-$\mathcal{O}(e_0^2)$ & $\mathcal{O}(e_0)$ & $(1,-2)$, $(2,-2)$,\newline $(3,-2)$ \\
\texttt{TaylorF2Ecc} & 3PN-$\mathcal{O}(e_0^2)$ & None & $(2,-2)$ \\
\texttt{TaylorF2} & 3, 3.5, 4, 4.5PN & None & $(2,-2)$ \\
\hline\hline
\end{tabular}
\label{table:approximants}
\end{table}

We now describe the version of the \texttt{TaylorF2Ecc} approximant that we employ, which is detailed in Sec.~IV of Ref.~\cite{MF16} and mentioned in Sec.~II of Ref.~\cite{Favata2022}. This approximant extends the quasicircular \texttt{TaylorF2} approximant by incorporating leading-order eccentric corrections, up to 3PN order, in the Fourier phase. Its amplitude is therefore circular and accurate to Newtonian (quadrupolar) order, restricting its harmonic content to the dominant $(2,-2)$. As detailed in Ref.~\cite{MF16}, \texttt{TaylorF2Ecc} also includes circular 3.5PN-order contributions to the Fourier phase.
To make meaningful comparisons in our analysis, we typically ignore the circular nonspinning 3.5PN terms and PN-accurate (circular) spin contributions that appear in the Fourier phase expression of the \texttt{TaylorF2Ecc} approximant, unless stated otherwise. This approximant does not account for periastron advance in either its Fourier phases or its amplitudes, as is evident from Eq.~(6.26) of Ref.~\cite{MF16}.

Finally, we briefly discuss the versions of the quasicircular \texttt{TaylorF2} approximant, detailed in Ref.~\cite{BIOPS}, that we employ in the present work. This approximant describes purely circular inspirals and uses a 3.5PN-accurate Fourier phase expression given by Eq.~(3.18) in Ref.~\cite{BIOPS}. The standard \prdsoftware{LALSuite} implementation of \texttt{TaylorF2} typically employs Newtonian (quadrupolar) amplitude corrections, with phase corrections accurate to 3.5PN order with respect to the dominant quadrupolar contribution. To probe systematics in the parameter-estimation studies of our eccentric approximants, we employ versions of \texttt{TaylorF2} whose Fourier phases are accurate to 3PN, 3.5PN, 4PN, and 4.5PN order. For the 4PN and 4.5PN versions, we use Eq.~(9) of Ref.~\cite{B45PN_23}, which provides a fully 4.5PN-accurate Fourier phase for circular inspirals. Table~\ref{table:approximants} lists all waveform approximants used in the present investigations.

\subsection{Waveform sanity checks}\label{sec:sanity_check}

\begin{figure*}[ht!]
    \centering
    \includegraphics{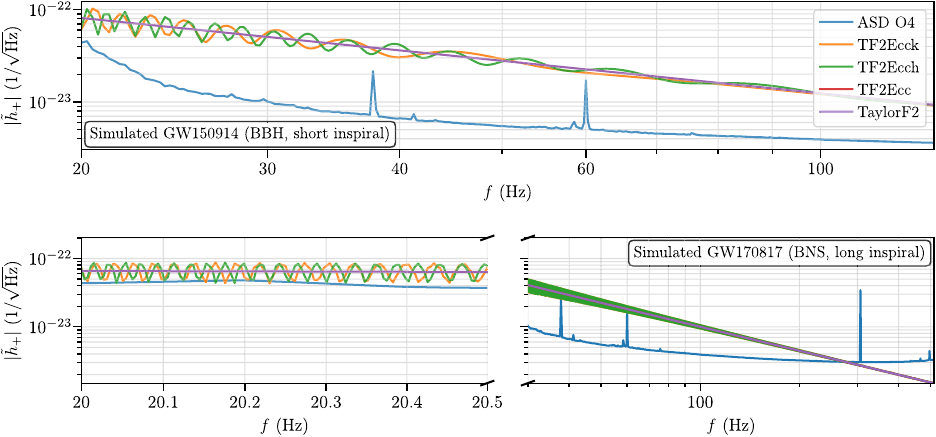}
    \caption{Plot illustrating the amplitude evolution ($|\tilde{h}_+(f)|$) as a function of frequency for eccentric waveform approximants (\texttt{TaylorF2Ecck}, \texttt{TaylorF2Ecch}) compared to \texttt{TaylorF2Ecc} and the quasicircular \texttt{TaylorF2} model. The top panel uses parameters similar to GW150914 (BBH, short inspiral), while the bottom panel uses parameters similar to GW170817 (BNS, long inspiral), both with an initial eccentricity $e_0 = 0.1$ at $f_0 = 20$~Hz. The amplitude spectral density (ASD) for the LIGO L1 detector at O4 design sensitivity is overplotted for reference. For this moderate, non-negligible value of $e_0$, both \texttt{TaylorF2Ecck} and \texttt{TaylorF2Ecch} exhibit distinct amplitude modulations due to multiple harmonics, with \texttt{TaylorF2Ecck} displaying a different modulation morphology resulting from additional sidebands (extra harmonics) and phase shifts in the subdominant modes. At higher frequencies, as eccentricity decays, the modulations diminish, and the waveforms gradually converge to the circular \texttt{TaylorF2} model.}\label{fig:sanity-check1}
\end{figure*}
Initial verifications of our \texttt{TaylorF2Ecck} eccentric inspiral approximant within the \texttt{LALSimulation} framework were performed by comparing it with \texttt{TaylorF2Ecch} and the well-established \texttt{TaylorF2} and \texttt{TaylorF2Ecc} inspiral template families. These checks focused on computational efficiency, waveform amplitude and phase evolution with frequency, and convergence to the quasicircular \texttt{TaylorF2} approximant as eccentricity approaches zero. Further checks were conducted to determine the range of parameter space for which the inclusion of eccentricity and periastron advance becomes important in waveform-level comparisons. Waveform deviations of \texttt{TaylorF2Ecck} from \texttt{TaylorF2}, \texttt{TaylorF2Ecc}, and \texttt{TaylorF2Ecch} were examined in the $m$--$e_0$ and $\eta$--$e_0$ parameter spaces. 
All analyses in this section use a minimum frequency of $20$~Hz, a choice discussed in detail in Sec.~\ref{sec:analysis_techniques}.
Additionally, we use GW170817-like parameters (specifically masses and distance) obtained from a parameter-estimation run with \texttt{TaylorF2Ecck} on the GW170817 data (more details about parameter estimation in Sec.~\ref{sec:pe_gw170817}).
It should be noted that \texttt{TaylorF2Ecch} is included in these tests for context only. Its waveform behavior will not be elaborated upon in the results of Sec.~\ref{sec:sanity_check}, and it will not be included in the event analysis of Sec.~\ref{sec:pe_gw170817}.

Given the GW source parameters, \texttt{LALSimulation} directly outputs the plus and cross polarizations, $\tilde{h}_+(f)$ and $\tilde{h}_{\times}(f)$, respectively. The corresponding detector strain is given by $\tilde{h}(f)=F_+\,\tilde{h}_+(f) + F_{\times}\,\tilde{h}_{\times}(f)$.
For simplicity, we use the plus-polarization modulus $|\tilde{h}_+(f)|$ to probe the amplitude evolution of our approximants. We have verified that our conclusions remain essentially unchanged if we instead use either the cross-polarization modulus $|\tilde{h}_{\times}(f)|$ or the full Fourier amplitude $|\tilde{h}(f)|$.
In Fig.~\ref{fig:sanity-check1}, we display $|\tilde{h}_+(f)|$ for the \texttt{TaylorF2Ecck}, \texttt{TaylorF2Ecch}, \texttt{TaylorF2Ecc}, and \texttt{TaylorF2} approximants, using simulated GW150914-like and GW170817-like parameters to illustrate their behavior in short- and long-inspiral scenarios. We take $e_0 = 0.1$ at $20$~Hz.

For $e_0 > 0.02$, both the \texttt{TaylorF2Ecck} and \texttt{TaylorF2Ecch} approximants exhibit visible amplitude modulations due to overlapping harmonic contributions; however, the morphology of these modulations differs because \texttt{TaylorF2Ecck} includes extra harmonics from sidebands and additional phase shifts in the subdominant modes. For low-total-mass systems, these approximants display a greater number of modulations over a given frequency range. As the eccentricity decreases at higher frequencies, these modulations diminish, and the waveforms converge toward their quasicircular \texttt{TaylorF2} counterpart (see Eq.~(\ref{eq:et}) in Appendix~\ref{sec:et_evolution} for the analytical formulation of eccentricity evolution with frequency). Additionally, the harmonic-dependent amplitude coefficients ($\xi_{j,n}$ and $\xi_{j}$) scale inversely with frequency through their dependence on the $\chi$ variable, further contributing to the decay of the amplitude modulations with frequency. In a supplementary check, to visualize the waveform oscillations and phase shifts in the frequency domain, we plot the real part of $\tilde{h}_+(f)$ (which scales with $\cos(\Psi_{j,n}-\pi/4)$). We find that it evolves similarly with frequency across the eccentric models while showing noticeable frequency-dependent phase shifts relative to the quasicircular \texttt{TaylorF2} model (see Fig.~\ref{fig:real_waveform_gw150914} in Appendix~\ref{sec:real_part_waveform_comparison}).

To assess the impact of eccentricity and periastron advance on waveform-level comparisons, we follow the convention used in the match study presented in Sec.~B of Ref.~\cite{TGHH} and employ match computations to probe the basic GW data-analysis implications of our inspiral templates. Following Ref.~\cite{Damour1998}, the match ${\cal M}(h_s,h_t)$ between a signal waveform $h_s$ and a template waveform $h_t$ is computed by maximizing the normalized overlap ${\cal O}(h_s,h_t)$ over the detector arrival time $t_0$ and phase $\phi_0$ of the template waveform. It reads
\begin{equation}
{\cal M}(h_s, h_t) = \max_{t_0, \phi_0} {\cal O}(h_s, h_t),
\label{eq:match_computation}
\end{equation}
\noindent where the normalized overlap is
\begin{equation}
{\cal O}(h_s,h_t)
=
\frac{(h_s|h_t)}
{\sqrt{(h_s|h_s)(h_t|h_t)}} \,.
\label{eq:normalized_overlap}
\end{equation}

The detector-noise-weighted inner product entering the overlap calculation reads
\begin{equation}
(h_s|h_t)
=
4\,{\rm Re}
\int_{f_{\rm min}}^{f_{\rm max}}
\frac{\tilde{h}^{*}_s(f)\tilde{h}_t(f)}
{S_h(f)}\,df .
\label{eq:noise_weighted_inner_product}
\end{equation}

Here, $\tilde{h}_s(f)$ and $\tilde{h}_t(f)$ are the Fourier transforms of the signal and template waveforms, and $S_h(f)$ denotes the one-sided detector power spectral density. For our match estimates, $S_h(f)$ is given by the LIGO O4 design sensitivity PSD~\cite{LIGO-T2000012} for the L1 detector. The integration limits are set to $f_{\rm min}=20$~Hz and $f_{\rm max}=f_{\rm LSO}$. The frequency resolution is $df=1/\Delta T$, where $\Delta T$ is the time-domain duration of the waveform. With the GW170817-like parameters used here, we obtain a signal-to-noise ratio of $\mathrm{SNR}\approx40$ in this setup. In practice, these match values are computed using the \texttt{match} routine from \texttt{pycbc.filter.matchedfilter}~\cite{pycbc}.

For our analysis, we use the mismatch percentage ${\cal M}^{*}_{\%} = (1-{\cal M}) \times 100\%$, rather than the match ${\cal M}$ itself~\cite{Khan2019,Wang2021}. A commonly used conventional limit in matched-filtering searches is ${\cal M}\geq0.97$, or equivalently ${\cal M}^{*}_{\%}\leq3\%$. This threshold corresponds to a loss of less than $10\%$ of signal events in matched-filter searches. Consequently, in regions of parameter space where the computed matches are high (i.e., ${\cal M}\geq0.97$), waveform models are generally considered effectual templates for detection~\cite{Damour1998}. In such searches, one is ultimately interested in the fitting factor, which is the overlap integral maximized over the whole parameter space explored by a template bank.

In the present work, however, we compute fixed-intrinsic-parameter matches across a range of $e_0$, $m$, and $\eta$ values. The signal and template waveforms are evaluated at the same intrinsic parameters, and the overlap is maximized only over the detector arrival time and phase. This provides a controlled local comparison between waveform families at physically corresponding source parameters and avoids additional mismatch from comparing different intrinsic points. For a continuous fitting-factor calculation in which the same intrinsic point is included in the allowed template parameter space, this fixed-parameter match is a lower bound on the fitting factor. However, it is not itself a fitting-factor calculation because we do not maximize over intrinsic source parameters or over a discrete template bank. Such a maximization could increase the match if shifts in the intrinsic parameters partially compensate for missing physical effects. We therefore treat the match as a search-oriented waveform-level diagnostic, similar in spirit to the estimates of Ref.~\cite{TGHH}.

However, a match value above the conventional search threshold does not by itself guarantee accurate parameter estimation. At high SNRs, even small waveform differences caused by missing physical effects, such as eccentricity or periastron advance, can be resolved by the data and may lead to systematic biases in the inferred source parameters. A sufficient condition for such waveform differences to be negligible is given by the indistinguishability criterion~\cite{creightonanderson2011}, i.e., $(h_s - h_t | h_s - h_t) < 1$. In terms of a critical match ${\cal M}_c$, this condition implies approximately $1-{\cal M}_c \lesssim 1/(2\rho^2)$, where $\rho$ denotes the SNR. This relation shows that waveform distinguishability depends on the signal amplitude, and that waveform differences may affect parameter estimation when the mismatch is larger than this SNR-dependent scale. In our analysis, we do not use this criterion as a threshold for parameter-estimation accuracy or waveform distinguishability, but we note its qualitative significance and its dependence on the SNR. Instead, we assess the impact of eccentricity and periastron advance on parameter estimation using injection-recovery studies and Bayes factor comparisons.

For the injection-recovery test, we consider an idealized scenario where a \texttt{TaylorF2Ecck} waveform is injected into zero noise. We then use \texttt{TaylorF2Ecck},
\par\onecolumngrid
\vfill
\noindent\begin{minipage}{\textwidth}
\centering
\includegraphics{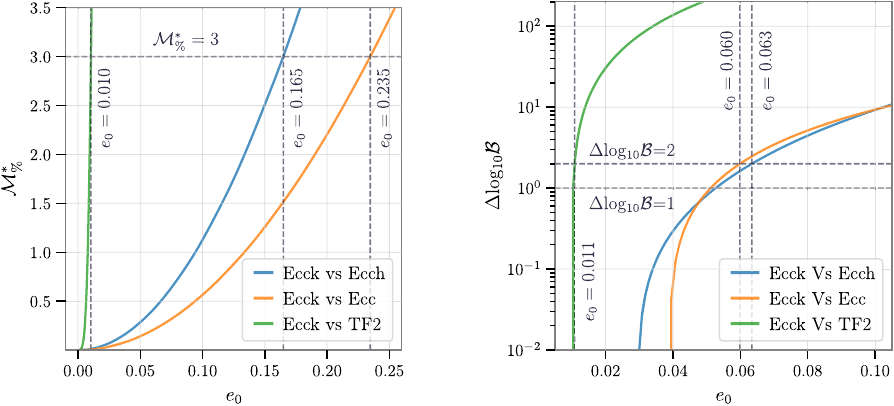}
    \captionof{figure}{Mismatch and Bayes-factor comparisons for the \texttt{TaylorF2Ecck} (Ecck) waveform model relative to eccentric inspiral approximants without periastron advance, \texttt{TaylorF2Ecch} (Ecch; blue) and \texttt{TaylorF2Ecc} (Ecc; orange), and the quasicircular \texttt{TaylorF2} (TF2; green) model.
    The {\it left panel} shows the mismatch percentage, $\mathcal{M}^{*}_{\%}$, as a function of $e_0$ for a single-detector setup with ${\rm SNR}\approx40$. Following the convention of Ref.~\cite{TGHH}, we mark $\mathcal{M}^{*}_{\%}=3$, or equivalently $\mathcal{M}=0.97$, with a horizontal dashed line as a search-oriented reference scale for waveform effectualness.
    As expected, in the limit $e_0 \to 0$, $\mathcal{M}^{*}_{\%}\to 0$, and all eccentric models tend toward the quasicircular model. In this setup, incorporating eccentricity becomes relevant (crossing the dashed line) at $e_0 \gtrsim 0.01$ when comparing \texttt{TaylorF2Ecck} with \texttt{TaylorF2}, while accounting for periastron advance becomes relevant at $e_0 \gtrsim 0.17$ when comparing with \texttt{TaylorF2Ecch}, and at $e_0 \gtrsim 0.24$ when comparing with \texttt{TaylorF2Ecc}.
    The {\it right panel} shows $\Delta\log_{10}{\cal B}$ as a function of $e_0$ from reduced five-dimensional injection-recovery analyses using a two-detector setup with network ${\rm SNR}\approx64$. The horizontal dashed lines at $\Delta\log_{10}{\cal B}=1$ and $2$ indicate strong and decisive evidence for waveform distinguishability, respectively. In this setup, decisive evidence is reached at $e_0 \simeq 0.01$ relative to \texttt{TaylorF2}, and at $e_0 \simeq 0.06$ relative to the eccentric models without periastron-advance effects, above which waveform differences can significantly affect parameter estimation. These quantitative thresholds are specific to the assumed source parameters, PSD, and network SNR.}\label{fig:sanity-check2}
\end{minipage}
\newpage
\twocolumngrid
\noindent
 \texttt{TaylorF2Ecch}, \texttt{TaylorF2Ecc}, and \texttt{TaylorF2} waveform templates to recover it via Bayesian parameter estimation, employing a reduced five-dimensional parameter space (i.e., $\mathcal{M}$, $q$, $e_0$, $\phi_c$, and $t_c$) instead of the standard 16 parameters~\cite{bilby_paper} to ensure reasonable computational cost. These injection-recovery tests use the LIGO L1 and H1 detectors with O4 design sensitivity PSDs and a minimum frequency of $20$~Hz. For the GW170817-like injection used in Fig.~\ref{fig:sanity-check2}, the injected signal has network ${\rm SNR}\approx64$ in this setup. For the supplementary GW190425-like injection discussed in Appendix~\ref{sec:appendix_injection_recovery}, the injected signal has network ${\rm SNR}\approx13$. 

Waveform distinguishability is assessed through differences in the resulting Bayes factors ($\Delta \log_{10} {\cal B}$), with values between $1$ and $2$ indicating strong evidence and values $>2$ indicating decisive distinguishability~\cite{Kass01061995}. 
We also note that alternative distinguishability criteria exist. For instance, a threshold of $|\Delta \log{\cal B}| = 8$ ($|\Delta \log_{10}{\cal B}| \approx 3.47$)~\cite{Thrane2019} is often quoted as ``strong evidence'' in favor of one hypothesis over another~\cite{Jeffreys1961}. We do not adopt this criterion in the present study.
Our injection-recovery framework further enables us to explore how \texttt{TaylorF2Ecck} deviates from other waveform families as the injected parameters are varied. In this context, the $\Delta \log_{10} {\cal B}=1$ (strong evidence) and $2$ (decisive evidence) thresholds serve as practical reference levels for identifying significant waveform differences. The threshold values quoted here should be interpreted only within the specific injection-recovery setup used in this work and are likely to change for different SNRs and PSDs. The technicalities of parameter estimation and Bayes-factor calculations are detailed in Sec.~\ref{sec:data_analysis}.

Figure~\ref{fig:sanity-check2} compares the \texttt{TaylorF2Ecck} approximant to other template families using match calculations and injection-recovery tests.
In the left panel, ${\cal M}^{*}_{\%}$ is plotted against $e_0$. The three colored curves represent match calculations of \texttt{TaylorF2Ecck} against \texttt{TaylorF2Ecch}, \texttt{TaylorF2Ecc}, and \texttt{TaylorF2}, respectively.
We employ ${\cal M}^{*}_{\%}$ rather than ${\cal M}$ to facilitate comparison with the distinguishability metric, $\Delta\log_{10}{\cal B}$.
Figure~\ref{fig:sanity-check2} (right panel) displays $\Delta\log_{10}{\cal B}$ values from the injection-recovery tests performed with varying injected values of $e_0$. The injected signals are generated using the \texttt{TaylorF2Ecck} inspiral waveform family with GW170817-like parameters. These signals are then recovered with the \texttt{TaylorF2Ecck}, \texttt{TaylorF2Ecch}, \texttt{TaylorF2Ecc}, and \texttt{TaylorF2} inspiral template families. For each model comparison, injections are performed at four discrete $e_0$ values between $0.0$ and $0.2$. The resulting discrete values of $\Delta\log_{10}{\cal B}$ are then interpolated using cubic splines.

We now discuss our inferences from the plots shown in Fig.~\ref{fig:sanity-check2}.
The quasicircular \texttt{TaylorF2} waveform diverges rapidly from \texttt{TaylorF2Ecck}, exceeding $3\%$ mismatch at $e_0 \approx 0.01$. At low eccentricities, all eccentric waveforms exhibit similar phase evolution. The \texttt{TaylorF2Ecch} and \texttt{TaylorF2Ecc} models reach $3\%$ mismatch at $e_0 \approx 0.17$ and $e_0 \approx 0.24$, respectively. For comparison, Ref.~\cite{TGHH} reports a threshold of $e_0 \sim 0.25$ at $20$~Hz from a comparison of eccentric models with and without periastron advance effects. The small difference is likely due to differences in the analysis settings, including the employed PSDs, the SNR used in Ref.~\cite{TGHH} ($\sim30$), and their use of beyond-leading-order eccentric corrections.
Injection-recovery tests indicate that \texttt{TaylorF2Ecck} becomes distinguishable from both \texttt{TaylorF2Ecch} and \texttt{TaylorF2Ecc} at $e_0 \gtrsim 0.05$ with strong evidence, and at $e_0 \gtrsim 0.06$ with decisive evidence. This provides insight into the $e_0$ parameter space where periastron advance significantly influences parameter estimation within this idealized setup. However, both \texttt{TaylorF2Ecc} and \texttt{TaylorF2Ecch} successfully recover the injected parameters within the $90\%$ credible intervals, even for larger injected $e_0$ values ($> 0.06$).
In contrast, \texttt{TaylorF2} failed to recover the injected mass parameters even for $e_0 \geq 0.01$, producing inaccurate parameter-estimation posteriors for this GW170817-like injection. Furthermore, \texttt{TaylorF2Ecck} becomes distinguishable, with decisive evidence, from \texttt{TaylorF2} at $e_0 \gtrsim 0.01$ in this setup, identifying a region in $e_0$ where eccentricity affects our reduced-dimensional injection-recovery analysis. Additional posterior plots of these analyses are included in Appendix~\ref{sec:appendix_injection_recovery_GW170817}.
We also perform supplementary sanity checks with GW190425-like parameters in Appendix~\ref{sec:appendix_injection_recovery_GW190425} to examine how the results change for a system with a different mass and SNR. This injection features a higher mass and, due to a larger luminosity distance, a lower network SNR (${\rm SNR}\approx13$) in the same O4-design L1--H1 setup. In this scenario, \texttt{TaylorF2Ecck} becomes distinguishable from the quasicircular \texttt{TaylorF2} waveform at $e_0 \gtrsim 0.018$ with decisive evidence, which is higher than the threshold obtained ($e_0 \gtrsim 0.011$) for the GW170817-like injection. This qualitatively reflects the SNR scaling of the indistinguishability criterion, $1-{\cal M}_c \lesssim 1/(2\rho^2)$, indicating that lower-SNR signals require a larger waveform mismatch, and consequently a higher eccentricity $e_0$, to become distinguishable. The differing mass, and consequently the differing signal duration, may also contribute to this shift in the $e_0$ threshold.

\begin{figure*}[ht!]
    \centering
    \includegraphics{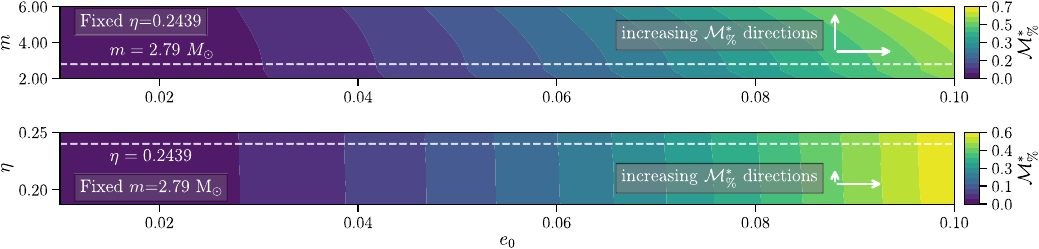}
    \caption{Mismatch percentage (${\cal M}^{*}_{\%}$) contours illustrating waveform deviations between the \texttt{TaylorF2Ecck} and \texttt{TaylorF2Ecc} approximants, probing the impact of periastron advance and additional harmonics. The top panel shows ${\cal M}^{*}_{\%}$ as a function of $m$ and $e_0$ (evaluated at $20$~Hz) with $\eta$ kept fixed. The bottom panel displays ${\cal M}^{*}_{\%}$ as a function of $\eta$ and $e_0$ with $m$ kept fixed. White dashed lines mark the parameters corresponding to a GW170817-like BNS system. While the waveform deviation generally increases with $e_0$, its dependence on the mass parameters varies. For tiny eccentricities ($e_0 \lesssim 0.02$), the mismatch is relatively insensitive to changes in $m$ and $\eta$. However, at larger values of $e_0$, ${\cal M}^{*}_{\%}$ increases noticeably with $m$ in the $e_0$--$m$ plane. A similar trend is observed in the $e_0$--$\eta$ plane, although the dependence on $\eta$ is relatively weaker over the considered parameter range. White arrows indicate the direction of increasing ${\cal M}^{*}_{\%}$, with their lengths qualitatively representing the local rate of change.}\label{fig:contour_mismatch}
\end{figure*}

To examine deviations between template families and their dependence on the source parameters, we conduct several match tests in the $e_0$--$m$ and $e_0$--$\eta$ parameter spaces. We consider ranges of $m$ and $\eta$ that are typical of BNS systems, with component masses in the range $1$--$3\,M_\odot$, and vary $e_0$ from low to moderate values of $0.01$--$0.10$. In the $e_0$--$m$ plane, only $e_0$ and $m$ are varied, while all other parameters are kept fixed, including $\eta$. Similarly, in the $e_0$--$\eta$ plane, only $e_0$ and $\eta$ are varied, while $m$ and all other parameters are kept fixed. The fixed parameters are chosen to be consistent with those used in our previous match tests. Performing similar investigations with injection-recovery tests would be prohibitively expensive. We therefore use variations in ${\cal M}^{*}_{\%}$ as a qualitative proxy for waveform deviations and for the trends expected in parameter-estimation studies. This use is motivated by Fig.~\ref{fig:sanity-check2}, where ${\cal M}^{*}_{\%}$ and $\Delta\log_{10}{\cal B}$ increase with $e_0$ in a similar qualitative way, although they are not quantitatively equivalent. A larger ${\cal M}^{*}_{\%}$ therefore indicates a larger waveform difference, which may suggest larger differences in parameter-estimation results, such as Bayes factors and posterior distributions.

Figure~\ref{fig:contour_mismatch} illustrates how ${\cal M}^{*}_{\%}$ varies with $m$, $e_0$, and $\eta$ when comparing \texttt{TaylorF2Ecck} with \texttt{TaylorF2Ecc}. This comparison probes the impact of including periastron advance and additional harmonics. An extended comparison involving \texttt{TaylorF2Ecch} and \texttt{TaylorF2} is provided in Appendix~\ref{sec:match_test_appendix}. While Fig.~\ref{fig:sanity-check2} established that the deviation between \texttt{TaylorF2Ecck} and \texttt{TaylorF2Ecc} increases with $e_0$, we now examine how this behavior is influenced by changes in $m$ and $\eta$.
In the $e_0$--$m$ plane, ${\cal M}^{*}_{\%}$ increases with $m$ at a fixed $e_0$ (see the top panel of Fig.~\ref{fig:contour_mismatch}). A similar trend is observed in the $e_0$--$\eta$ plane, although the dependence on $\eta$ is weaker over the considered parameter range (see the bottom panel of Fig.~\ref{fig:contour_mismatch}). For tiny eccentricities ($e_0 \lesssim 0.02$), the mismatch is relatively insensitive to changes in $m$ and $\eta$. However, at larger values of $e_0$, the dependence on these mass parameters becomes much more pronounced, particularly for $m$. This suggests that at low-to-moderate eccentricities, higher-mass BNS systems may exhibit larger waveform-level differences between \texttt{TaylorF2Ecck} and \texttt{TaylorF2Ecc}. Consequently, their parameter-estimation results may be more sensitive to the inclusion of periastron advance. Such behavior is expected because periastron advance is a post-Newtonian effect. Its contribution to the waveform becomes more pronounced for higher-mass systems, which generate stronger gravitational fields and thus experience more significant relativistic effects. A rigorous investigation of how this parameter dependence propagates into full parameter-estimation results is beyond the scope of this work and will be explored in future studies with an extended \texttt{TaylorF2Ecck} approximant.

To conclude, \texttt{TaylorF2Ecck} diverges from \texttt{TaylorF2} even at small eccentricities (i.e. $e_0\gtrsim 0.01$) in our idealized GW170817-like injection-recovery setup, where eccentricity corrections are important.
The $e_0$ limits from Ref.~\cite{Favata2022} support a similar conclusion, indicating significant systematic bias when eccentricity is ignored for $e_0\gtrsim 0.01$ at $10$~Hz.
Moreover, the periastron effect becomes important at comparatively small eccentricities in our setup ($e_0\gtrsim 0.06$). The quoted eccentricity threshold values are not universal thresholds. They depend on the assumed detector network, PSD, SNR, source parameters, and parameter space used in the recovery.
The real-event analyses of GW170817 and GW190425 in Sec.~\ref{sec:pe_gw170817} are therefore interpreted directly through their own Bayes factors and posterior distributions, rather than by adopting the eccentricity thresholds inferred from this sanity-check section.
Lastly, the sanity tests in this section indicate that \texttt{TaylorF2Ecck} can diverge from \texttt{TaylorF2Ecc} in the eccentricity--mass parameter space. The deviation is driven mainly by changes in $e_0$, but it also increases with $m$ and, more weakly, with $\eta$, especially at larger eccentricities.

\section{Bayesian parameter estimation studies on inspiral-dominated events}\label{sec:data_analysis}

This section presents Bayesian parameter-estimation studies for two inspiral-dominated GW events. We first provide a brief overview of the Bayesian methodology, which is widely used within the LVK collaboration for GW data analysis~\cite{creightonanderson2011, Sivia2006, bilby_paper}. This is followed by a detailed description of our parameter-estimation studies for these events, where we compare our eccentric inspiral templates with existing waveform families and explore their astrophysical and data-analysis implications.

\subsection{Bayesian inference framework}\label{sec:analysis_techniques}

Bayesian inference provides a robust framework for estimating the source and orbital parameters of GW events. This approach involves modeling the observed data, $d$, with an inspiral template family under a chosen hypothesis, $H$, assuming stationary Gaussian detector noise. The posterior probability density, ${\cal P}(\theta | d, H)$, for the source parameters $\theta$ given the data and the model is obtained via Bayes' theorem as
\begin{equation}
{\cal P}(\theta | d,H) = \, \frac{{\cal L}(d | \theta,H) \, \pi(\theta | H)}{{\cal Z}(d | H)} \,, \label{eq:bayes_theorem}
\end{equation}
\noindent where ${\cal L}(d | \theta, H)$ is the likelihood function, $\pi(\theta | H)$ is the prior distribution for the parameters, and ${\cal Z}(d | H)$ is the Bayesian evidence given by
\begin{equation}
{\cal Z}(d | H) = \int {\cal L}(d | \theta,H) \, \pi(\theta | H) \, d\theta \,. \label{eq:evidence}
\end{equation}

The likelihood function quantifies the agreement between the data and the model for a given set of parameters, weighted against the Gaussian noise profile. For our analysis, we employ the \prdsoftware{Bilby} \prdsoftware{Python} package~\cite{bilby_paper, bilby_pipe_paper}, which implements both nested sampling and Markov chain Monte Carlo (MCMC) algorithms for posterior computation.
Nested sampling, in contrast to the MCMC approach used in Ref.~\cite{LNB}, is particularly advantageous for waveform model comparison, as it provides an estimate of the Bayesian evidence, enabling the calculation of the Bayes factor. For a given parameter-estimation run, \prdsoftware{Bilby} outputs the base-10 logarithm of the Bayes factor for the signal-versus-noise hypothesis, $\log_{10}{\cal B} = \log_{10}({\cal Z}/{\cal Z}_{\rm noise})$. To evaluate the relative evidence for model A over model B, we compute the logarithmic relative Bayes factor $\Delta \log_{10}{\cal B} \equiv \log_{10}{\cal B}_B^A = \log_{10}({\cal Z}_A/{\cal Z}_B)$. Because the noise evidence ${\cal Z}_{\rm noise}$ is identical for both runs, this metric is straightforwardly obtained by taking the difference of their respective \prdsoftware{Bilby} outputs. Throughout our analysis, waveform comparisons rely on this logarithmic metric, as previously employed in the injection-recovery tests of Sec.~\ref{sec:sanity_check}. Specifically for event analysis, if $|\Delta \log_{10}{\cal B}| < 1$, we conclude there is no strong evidence favoring one waveform model over the other. We also expect that, if the posterior favors negligible eccentricity, all eccentric waveforms become indistinguishable from each other and from the quasicircular model, resulting in a small magnitude for $|\Delta \log_{10}{\cal B}|$.

Our eccentric inspiral template families are implemented in \texttt{LALSimulation}, which is part of \prdsoftware{LALSuite} and is written primarily in C for computational efficiency. \prdsoftware{Bilby} is then used to interface with \texttt{LALSimulation} for parameter estimation. Detailed settings and all parameter-estimation results for this work are archived in Ref.~\cite{TaylorF2EcckRepo}. We neglect tidal effects in our parameter-estimation runs and set priors similarly to Ref.~\cite{LNB}. Finally, we set all spin parameters to zero, unless stated otherwise.

The choice of analysis duration and $f_{\rm min}$ significantly affects both the accuracy and computational cost of our parameter estimation. To capture eccentricity effects, the maximum feasible data length, as fixed by the mass prior, should be used. For mass priors with $m_1,\,m_2\in [1~M_{\odot},\,3~M_{\odot}]$, the maximum data length corresponds to $\approx 309$~s (1.1 times the chirp time~\cite{lalsuite}) for a $1~M_{\odot}$--$1~M_{\odot}$ system, and it exceeds typical LVK analysis durations for BNS events (128--190~s)~\cite{GW170817pe,LNB}. Although lowering $f_{\rm min}$ could improve eccentricity measurements~\cite{Favata2022, gupte2024}, it drastically increases computational demands. For instance, decreasing $f_{\rm min}$ from $20$~Hz to $10$~Hz for GW170817 would require a sixfold increase in data duration.
Furthermore, the detector noise at $10$~Hz is significantly higher than at $20$~Hz~\cite{Sun2020, Sun2021}, which can introduce additional uncertainty. Given these factors, and the recommendation of Ref.~\cite{emma2024} that $20$~Hz is optimal for parameter estimation, we set $f_{\rm min}$ to $20$~Hz, acknowledging the potential loss of information below this frequency. To manage computational cost and complete a parameter-estimation run within about one week, we also set the sampling frequency to $2048$~Hz and $f_{\rm max}$ to 256~Hz for GW170817. This choice is justified because our primary interest lies in the inspiral phase, particularly the early inspiral where signatures of residual eccentricity, if present, are expected to be most pronounced. This approach is further supported by similar strategies in recent literature, such as the use of $f_{\rm max} = 280$~Hz to investigate eccentricity in the inspiral phase of an NSBH system~\cite{Morras2025b}.

For GW170817 and GW190425 parameter-estimation runs, we use the official \texttt{BayesWave}~\cite{bayeswave} PSDs provided in the data releases~\cite{LIGO-P1800061, LIGO2019_GW190425_data, LIGO-P2000026}, ensuring consistency with the original LVK analyses~\cite{GW170817pe, Abbott2020GW190425}. Supplementary parameter-estimation runs using PSDs computed via Welch's method (the \prdsoftware{Bilby} default) confirm that our conclusions remain unaffected by this choice.

\subsection{Parameter estimation implications of the \texttt{TaylorF2Ecck} approximant for inspiral-dominated events GW170817 and GW190425}\label{sec:pe_gw170817}

\begin{figure*}[ht!]
    \centering
    \includegraphics{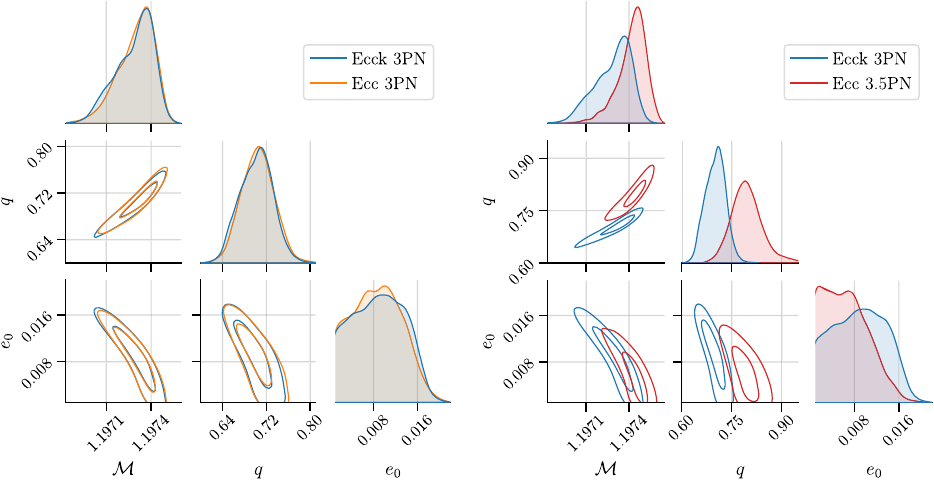}
    \caption{Corner plots showing the posterior distributions of ${\cal M}$, $q$, and $e_0$ from parameter estimation for GW170817 with a uniform $e_0$ prior. The 2D contours enclose 39.3\% and 86.4\% of the posterior samples.
    The {\it left panel} compares two nonspinning, 3PN-accurate eccentric inspiral waveform models, \texttt{TaylorF2Ecck} (Ecck 3PN) and \texttt{TaylorF2Ecc} (Ecc 3PN). While this panel provides our primary comparison, the {\it right panel} offers additional context by comparing \texttt{TaylorF2Ecck} with the \prdsoftware{LALSuite}-standard, but nonspinning, \texttt{TaylorF2Ecc} (Ecc 3.5PN), which includes 3.5PN-order contributions to the circular part of the Fourier phase.
    In the left panel, both 3PN models constrain the residual eccentricity to be negligible ($e_0\leq0.016$ for \texttt{TaylorF2Ecck} and $e_0\leq0.015$ for \texttt{TaylorF2Ecc}), and the $e_0$ posteriors rail towards zero. For details on the 90\% credible intervals and Bayes-factor comparisons, see Table~\ref{tab:e0_waveform_comparison}. At these negligible eccentricities, joint posteriors involving mass and eccentricity (e.g., the ${\cal M}$--$e_0$ and $q$--$e_0$ planes) and the one-dimensional marginalized distributions show no significant differences between the two 3PN models. This indicates that the inclusion of additional harmonics and periastron advance in \texttt{TaylorF2Ecck} has a negligible impact on parameter estimation for GW170817 in this regime. However, the right panel shows a substantial bias in the $q$ posterior when comparing the 3PN models with the 3.5PN \texttt{TaylorF2Ecc}, highlighting the influence of higher-order circular Fourier phase corrections.}\label{fig:pe_gw170817}
\end{figure*}
We now pursue parameter-estimation studies of our eccentric approximants using data from inspiral-dominated events. This effort primarily investigates the data-analysis implications of \texttt{TaylorF2Ecck} and compares them with parameter-estimation results obtained using the \texttt{TaylorF2Ecc} approximant, as reported in Refs.~\cite{LNB, Favata2022}.
Parameter estimation with the \texttt{TaylorF2Ecck} approximant poses significant computational challenges due to its complex harmonic structure and the prolonged duration of BNS signals. For instance, our Bayesian parameter estimation runs for GW170817 using \texttt{TaylorF2Ecck} took more than a week to complete, whereas the same analysis with the \texttt{TaylorF2Ecc} approximant required roughly half as much time. To mitigate this computational burden, we adopted the prior outlined in Ref.~\cite{LNB}, which uses electromagnetic counterpart data for GW170817 to fix the sky localization and constrain the luminosity distance prior to a Gaussian distribution centered on $40.7$~Mpc. However, this approach is not feasible for the GW190425 event, which lacks an electromagnetic counterpart and has a lower network ${\rm SNR} \approx 13$ compared to GW170817's ${\rm SNR} \approx 33$. This lower SNR complicates parameter constraints for GW190425, yet its inclusion in our study shows that key findings from GW170817 extend to other BNS inspiral events. 

While a detailed analysis of spin effects is beyond the scope of this work, Fig.~\ref{fig:pe_gw170817_compare} in Appendix~\ref{sec:full_gw170817} provides a broader context. Specifically, it compares posteriors derived from our 3PN-accurate \texttt{TaylorF2Ecck} approximant against those from the standard full \texttt{TaylorF2Ecc} model (which incorporates leading-order spin effects and 3.5PN circular contributions) using our own parameter-estimation setup. For external comparison, the figure also includes results from the same \texttt{TaylorF2Ecc} model analyzed by Lenon \textit{et al.}~\cite{LNB} under different settings, as well as the quasicircular \texttt{IMRPhenomPv2NRT} approximant employed by the LVK consortium.

\begin{figure}[t!]
\centering
\includegraphics{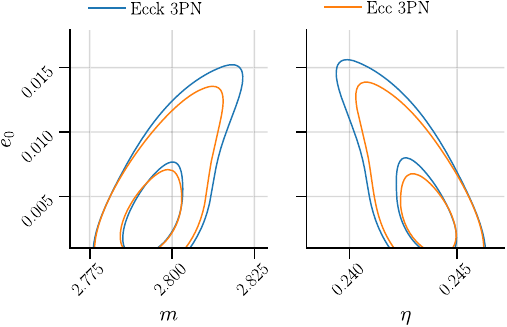}
    \caption{Contour plots for $m$ versus $e_0$ ({\it left panel}) and $\eta$ versus $e_0$ ({\it right panel}), comparing \texttt{TaylorF2Ecck} (Ecck, blue) and \texttt{TaylorF2Ecc} (Ecc, orange), from parameter estimation for GW170817 with a log-uniform $e_0$ prior (see Fig.~\ref{fig:GW170817_full_pe} for results with a uniform prior). The contours enclose $39.3\%$ and $86.4\%$ of the posterior mass. There is a positive correlation between $m$ and $e_0$ and a negative correlation between $\eta$ and $e_0$. At these negligible eccentricities, we observe subtle differences between the two approximants at higher $e_0$ values. However, these differences are even less pronounced under a uniform $e_0$ prior and should be interpreted with caution.}\label{fig:correlation_gw170817}
\end{figure}

Figure~\ref{fig:pe_gw170817} presents a comparative analysis of the posterior distributions for ${\cal M}$, $q$, and $e_0$ obtained from GW170817 parameter-estimation runs using the \texttt{TaylorF2Ecck} and \texttt{TaylorF2Ecc} waveform approximants. Complementary to this, Fig.~\ref{fig:correlation_gw170817} illustrates degeneracies and correlations in the joint posterior distributions across the $m$--$e_0$ and $\eta$--$e_0$ planes. Our primary comparison in Fig.~\ref{fig:pe_gw170817} (left panel) focuses on nonspinning, 3PN-accurate eccentric waveform models, while Fig.~\ref{fig:pe_gw170817} (right panel) includes the 3.5PN version of \texttt{TaylorF2Ecc}, with higher-order corrections applied only to the circular part of the Fourier phase, to highlight potential biases introduced by such terms. Initial observations from Fig.~\ref{fig:pe_gw170817} (left panel) suggest that, at the negligible eccentricities inferred for GW170817, the \texttt{TaylorF2Ecck} and \texttt{TaylorF2Ecc} models yield broadly consistent parameter constraints. However, a notable bias emerges in the $q$ posterior when comparing the 3PN \texttt{TaylorF2Ecck} to the 3.5PN-corrected \texttt{TaylorF2Ecc}, with the latter displaying a distinct shift in the posterior peak. This finding requires further exploration because eccentric contributions are currently fully known only up to 3PN order.

For GW170817, the $e_0$ posteriors for both \texttt{TaylorF2\-Ecck} and \texttt{TaylorF2Ecc} favor negligible eccentricity and rail toward zero. Using a uniform (log-uniform) prior on $e_0$, the 90\% credible upper limits are $e_0\leq0.016$ ($e_0\leq0.015$) for \texttt{TaylorF2Ecck} and $e_0\leq0.015$ ($e_0\leq0.013$) for \texttt{TaylorF2Ecc} (see Table~\ref{tab:e0_waveform_comparison} and Fig.~\ref{fig:pe_gw170817}). Employing the full \texttt{TaylorF2Ecc} approximant with a uniform prior on $e_0$, Lenon \textit{et al.}~\cite{LNB} report an upper limit of $e_0 \leq 0.024$ at $10$~Hz and their supplementary data show a closely matching limit of $e_0 \leq 0.013$ at $20$~Hz. In our analysis, Bayes factors show no clear preference for the eccentric models over the quasicircular \texttt{TaylorF2} model ($|\Delta \log_{10}{\cal B}| < 1$), which is consistent with negligible eccentricity in GW170817. Additionally, the Bayes-factor comparison between \texttt{TaylorF2Ecck} and \texttt{TaylorF2Ecc} gives no strong preference for either model ($|\Delta \log_{10}{\cal B}| < 1$). This is expected, as both eccentric models reduce to the quasicircular limit at low eccentricities, rendering the impact of periastron advance and extra harmonics negligible in this regime. Finally, we note that Lenon \textit{et al.}~\cite{LNB} do not report Bayes factors for direct comparison.

Figure~\ref{fig:correlation_gw170817} compares the \texttt{TaylorF2Ecck} and \texttt{TaylorF2Ecc} posteriors in the $m$--$e_0$ and $\eta$--$e_0$ planes for GW170817, highlighting the degeneracies between eccentricity and the mass parameters. We observe a positive correlation between $m$ and $e_0$ and a negative correlation between $\eta$ and $e_0$. These trends are consistent with our small-eccentricity injection-recovery tests (see Fig.~\ref{fig:injection_recovery1} in Appendix~\ref{sec:appendix_injection_recovery}). Interestingly, these correlations are opposite in sign to those reported in studies employing the full \texttt{TaylorF2Ecc} approximant~\cite{Favata2022, LNB}. This difference appears to be associated with the inclusion of leading-order spin effects in the Fourier phasing, rather than with the quasicircular 3.5PN Fourier phasing corrections (see Figs.~\ref{fig:GW170817_full_pe} and~\ref{fig:pe_gw170817_compare}). The two waveform models also show slightly different posterior morphologies. This is most visible in the log-uniform $e_0$ prior runs, where the two-dimensional contours in the $m$--$e_0$ and $\eta$--$e_0$ planes differ, especially at higher values of $e_0$ (see Fig.~\ref{fig:correlation_gw170817}). The difference is less visible for the uniform $e_0$ prior (see Fig.~\ref{fig:GW170817_full_pe}). Since GW170817 favors negligible eccentricity, these differences should be interpreted with caution, and we do not expect large waveform-dependent posterior shifts in this regime. This is consistent with the injection-recovery tests in Sec.~\ref{sec:sanity_check}, although those tests were performed at a higher SNR. We note that the posterior contours, when comparing the two eccentric models, can become more visibly different when noisier PSDs are used, such as PSDs computed via Welch's method using default \prdsoftware{Bilby} settings. Even in such cases, the inferred eccentricity remains negligible, and the Bayes-factor differences remain small, with $|\Delta \log_{10}{\cal B}|<1$.

Finally, the posteriors for luminosity distance ($d_L$) and inclination angle ($\theta_{jn}$) are remarkably similar between \texttt{TaylorF2Ecck} and \texttt{TaylorF2Ecc}. Unlike the mass parameters, we find no significant correlation between $e_0$ and either $d_L$ or $\theta_{jn}$. The full parameter posteriors for GW170817 are provided in Fig.~\ref{fig:GW170817_full_pe} of Appendix~\ref{sec:full_gw170817} for reference.

A similar analysis of GW190425 reinforces the conclusions drawn from GW170817, although with greater uncertainty due to its lower SNR and the lack of an electromagnetic counterpart. This leads to weaker constraints on the mass ratio and the eccentricity. While the inclusion of 3.5PN Fourier phase corrections in \texttt{TaylorF2Ecc} still induces a shift in the mass ratio, this effect is less pronounced than in GW170817. The full parameter posteriors are provided in Fig.~\ref{fig:GW190425_full_pe} of Appendix~\ref{sec:full_GW190425} for reference.
\begingroup
\renewcommand{\arraystretch}{1.5}
\begin{table}[t!]
\centering
\setlength{\abovecaptionskip}{0pt}
\caption{Summary of parameter-estimation results for GW170817 and GW190425 using the \texttt{TaylorF2Ecck} (\texttt{Ecck}) and \texttt{TaylorF2Ecc} (\texttt{Ecc}) waveform models. We report the median initial eccentricity $e_0$ and its 90\% credible intervals, evaluated at a reference frequency of $20$~Hz, under both uniform and log-uniform $e_0$ priors. The final column lists the corresponding $\log_{10}{\cal B}$ for the signal-versus-noise hypothesis. The quasicircular \texttt{TaylorF2} (\texttt{TF2}) model is included as a baseline reference. As shown, the constraints on $e_0$ are consistent with negligible eccentricity (see the $e_0$ posteriors for GW170817 in Fig.~\ref{fig:GW170817_full_pe} and GW190425 in Fig.~\ref{fig:GW190425_full_pe}, which rail towards zero). Furthermore, the relative Bayes factors $\Delta \log_{10}{\cal B}$ among the waveform models are small ($|\Delta \log_{10}{\cal B}| < 1$), indicating no strong preference for any specific model. Thus, at these negligible eccentricities, the inclusion of eccentricity and effects arising from periastron advance and multiple harmonics has no significant impact on parameter estimation for these two inspiral-dominated events.}
\small
\renewcommand{\arraystretch}{1.2}
\setlength{\tabcolsep}{2pt}
\centering
\begin{tabular*}{\columnwidth}{@{\extracolsep{\fill}}lllcc@{}}
\hline\hline
Event     & $e_0$ prior       & Model & $e_0$ value & $\log_{10}{\cal B}$  \\
\hline
GW170817  & Uniform     & \texttt{Ecck}  & $0.009^{+0.007}_{-0.008}$ & 220.24 \\
          &             & \texttt{Ecc}   & $0.008^{+0.007}_{-0.007}$ & 220.21 \\
          & Log-uniform & \texttt{Ecck}  & $0.005^{+0.010}_{-0.003}$ & 220.84 \\
          &             & \texttt{Ecc}   & $0.004^{+0.009}_{-0.003}$ & 220.76 \\
          & $\cdots$    & \texttt{TF2}   & $\cdots$                   & 220.71 \\
\hline
GW190425  & Uniform     & \texttt{Ecck}  & $0.010^{+0.013}_{-0.009}$ & 21.50 \\
          &             & \texttt{Ecc}   & $0.010^{+0.013}_{-0.009}$ & 21.36 \\
          & Log-uniform & \texttt{Ecck}  & $0.004^{+0.014}_{-0.003}$ & 21.91 \\
          &             & \texttt{Ecc}   & $0.004^{+0.014}_{-0.003}$ & 21.89 \\
          & $\cdots$    & \texttt{TF2}   & $\cdots$                   & 21.50 \\
\hline\hline
\end{tabular*}
\label{tab:e0_waveform_comparison}
\end{table}
\endgroup

\begin{figure}[t!]
\centering
\includegraphics{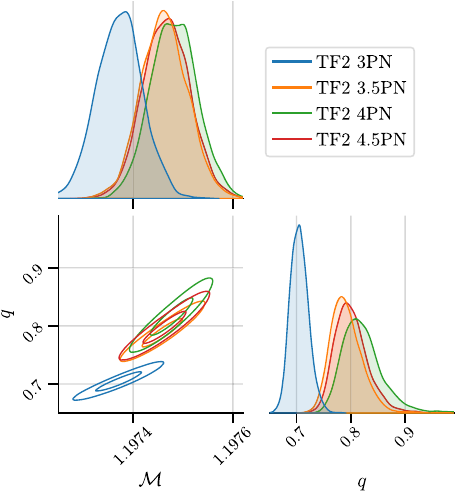}
    \caption{Chirp mass ${\cal M}$ and mass ratio $q$ posteriors for GW170817 evaluated using the quasicircular \texttt{TaylorF2} approximant (TF2) truncated at 3PN, 3.5PN, 4PN, and 4.5PN orders.
     These posteriors largely converge beyond 3PN when comparing medians and 90\% credible intervals.
     The implication is that inspiral approximants such as \texttt{TaylorF2Ecck} and \texttt{TaylorF2Ecc} should include Fourier phase corrections at least through 3.5PN order to yield reliable posteriors for these parameters.} 
\label{fig:TF2GW170817}
\end{figure}

For GW190425, much like GW170817, both \texttt{TaylorF2Ecck} and \texttt{TaylorF2Ecc} favor negligible eccentricity. The $e_0$ posteriors rail toward zero, and the Bayes-factor comparison with the quasicircular \texttt{TaylorF2} model shows no clear preference for either model, with $|\Delta \log_{10}{\cal B}| < 1$. Using a uniform (log-uniform) prior on $e_0$, the 90\% credible upper limits are $e_0\leq0.023$ ($e_0\leq0.018$) for \texttt{TaylorF2Ecck} and $e_0\leq0.023$ ($e_0\leq0.018$) for \texttt{TaylorF2Ecc} (see Table~\ref{tab:e0_waveform_comparison} and Fig.~\ref{fig:GW190425_full_pe}). We also find no significant evidence for the effects of periastron advance or extra harmonics, since the comparison between \texttt{TaylorF2Ecck} and \texttt{TaylorF2Ecc} gives $|\Delta \log_{10}{\cal B}| < 1$. Prior work by Lenon \textit{et al.}~\cite{LNB}, using the full \texttt{TaylorF2Ecc} model, reports 90\% credible upper limits of $e_0 \leq 0.048$ ($e_0 \leq 0.023$) at $10$~Hz under a uniform (log-uniform) prior. At $20$~Hz, their supplementary data show $e_0 \leq 0.025$ ($e_0 \leq 0.016$) under the same priors, which is highly consistent with our own 90\% credible upper limits.

In agreement with our findings for GW170817, both eccentric models exhibit highly similar mass-eccentricity correlations for GW190425. The $m$--$e_0$ and $\eta$--$e_0$ contours show no visible difference between \texttt{TaylorF2Ecck} and \texttt{TaylorF2Ecc} in this low-eccentricity regime (see Fig.~\ref{fig:GW190425_full_pe} of Appendix~\ref{sec:full_GW190425}). As with GW170817, we also find no visible correlation between $e_0$ and either $d_L$ or $\theta_{jn}$ for this event.

To further investigate the observed mass-ratio shift between the 3PN and 3.5PN waveform models, we performed additional checks with the quasicircular \texttt{TaylorF2} approximant at several PN orders in the Fourier phase. This supplementary analysis helps isolate how higher-order Fourier phase corrections affect parameter estimation.

\subsection{Probing the PN-order dependence of ${\cal M}$ and $q$ posteriors for GW170817}\label{sec:tf2_analysis}

The observed dependence of the median $q$ values on the PN order to which Fourier phase contributions are included, as evident from Fig.~\ref{fig:pe_gw170817} (right panel), prompted us to ask whether similar behavior appears for non-eccentric \texttt{TaylorF2} templates. In other words, do GW170817 posteriors from the quasicircular \texttt{TaylorF2} approximant depend on the PN accuracy of its Fourier phase?

The \texttt{TaylorF2} approximant is typically used with 3.5PN-accurate Fourier phases, as in Ref.~\cite{BIOPS}. It can be extended so that the Fourier phase is accurate to 4.5PN order~\cite{B45PN_23}. This extension is primarily due to a highly detailed and demanding effort that extended 
GW generation to 4.5PN order by employing the Multipolar-Post-Minkowskian approach (see Ref.~\cite{blanchet2024, B45PN_23} and references therein). More importantly, the relevant Fourier coefficients at all PN orders up to 4.5PN were provided in Ref.~\cite{B45PN_23}. Using these inputs, we extended the standard \texttt{TaylorF2} implementation in \prdsoftware{LALSuite} to 4.5PN for parameter estimation, yielding template variants with Fourier phases truncated at 3PN, 3.5PN, 4PN, and 4.5PN, which we denote as TF2 3PN, TF2 3.5PN, TF2 4PN, and TF2 4.5PN, respectively.

The results are shown in Fig.~\ref{fig:TF2GW170817}. For these runs, we use a sampling frequency of $2048$~Hz with $f_{\rm max}=1024$~Hz, while all other parameter-estimation settings remain the same as those in Sec.~\ref{sec:pe_gw170817}. Varying $f_{\rm max}$ does not alter our conclusions. The median values of both ${\cal M}$ and $q$ for TF2 3PN differ substantially from their higher-order counterparts. Notably for $q$, the posterior overlap between TF2 3PN and the other variants is minimal, while the overlaps among TF2 3.5PN, TF2 4PN, and TF2 4.5PN are large. Beyond 3.5PN, the ${\cal M}$ and $q$ posteriors exhibit only a weak dependence on additional PN corrections. This is reassuring, as our results indicate that the widely quoted ${\cal M}$ and $q$ posteriors for GW170817~\cite{GW170817pe} are indeed robust.

A natural corollary of these parameter-estimation studies is that it is desirable to extend the work of Ref.~\cite{Arun09}, which computed the angular momentum flux for eccentric binaries at 3PN order, to 3.5PN accuracy. This will involve computing 3.5PN-accurate mass and current quadrupole moments, which are required to obtain 3.5PN-accurate energy and angular momentum fluxes that mainly involve hereditary contractions~\cite{Arun09,LB_LR}. After that, it should be straightforward to obtain $e_t(x,x_0,e_0)$ and $\Psi(x, x_0, e_0)$ to 3.5PN order following the prescription in Ref.~\cite{TGHH}. Interestingly, this extension would not require inputs from the 4PN-accurate Keplerian-type parametric solution (for eccentric, nonspinning binaries) available in Ref.~\cite{CTGL21}, or the 4.5PN phasing if it becomes available later, to achieve reliable posterior results in parameter estimation.

In our opinion, it is important to obtain a \texttt{TaylorF2Ecck} approximant whose Fourier-phase coefficients are accurate to 3.5PN order for data-analysis purposes and while trying to obtain eccentric IMR template families that are required to search for eccentric GW events in LVK data. Additionally, it will be interesting to explore the implications of missing 3.5PN-order contributions to GW emission in the time-domain eccentric IMR template families that are being constructed these days~\cite{Huerta2018}.

\section{Conclusion}\label{sec:conclusion}

In this work, we introduced and tested \texttt{TaylorF2Ecck}, a \texttt{TaylorF2}-family frequency-domain inspiral approximant implemented in the \prdsoftware{LALSuite} framework. This model, influenced by Ref.~\cite{TGHH}, analytically describes gravitational waves from nonspinning compact binaries and consistently incorporates orbital, advance of periastron, and gravitational-wave emission effects fully up to 3PN order.
Specifically, \texttt{TaylorF2Ecck} incorporates ${\cal O}(e_0^2)$ contributions to the 3PN-accurate Fourier phases, with quadrupolar-order amplitudes accurate to ${\cal O}(e_0)$.

Following comprehensive sanity checks to validate the model's performance and investigate the influence of eccentricity and periastron advance in the relevant parameter space, we employed \texttt{TaylorF2Ecck} using the \prdsoftware{Bilby} software to analyze public LVK data for the inspiral-dominated events GW170817 and GW190425. Our Bayesian parameter-estimation results show that the inferred values of $e_0$ for both events are negligible, with the posterior distributions railing toward zero under both uniform and log-uniform priors on $e_0$. Specifically, we obtain 90\% credible upper limits of $e_0 \leq 0.016$ for GW170817 and $e_0 \leq 0.023$ for GW190425 at $20$~Hz, in line with Ref.~\cite{LNB}. Supporting this, our Bayes factor comparisons show no preference for the eccentric models over the quasicircular model ($|\Delta \log_{10}{\cal B}| < 1$), consistent with the expected near-degeneracy of eccentric and quasicircular templates at very low $e_0$. Furthermore, at the negligible eccentricities inferred for these events, comparisons between \texttt{TaylorF2Ecck} and \texttt{TaylorF2Ecc} also yield $|\Delta \log_{10}{\cal B}| < 1$, indicating no strong preference for the eccentric model that includes periastron advance with additional harmonics over the model without these features. Consequently, in this regime, these additional physical effects have no significant impact on parameter estimation, resulting in no discernible difference in the two-dimensional mass--eccentricity posterior contours. However, a substantial bias in the $q$ posterior is observed when comparing these models to a version of \texttt{TaylorF2Ecc} that includes 3.5PN-order circular contributions in its Fourier phase. Considering the known correlation between mass and eccentricity, these parameter-estimation studies suggest the importance of incorporating $e_0$ contributions up to at least 3.5PN order in our eccentric waveform model. This inference regarding the $q$ bias and the significance of 3.5PN Fourier phase corrections is further substantiated by our analysis using versions of the quasicircular \texttt{TaylorF2} approximant that incorporate Fourier phase contributions at various PN orders.

We can directly compare our $e_0$ constraints for GW190425, derived using \texttt{TaylorF2Ecck}, with those obtained by Ref.~\cite{LNB} using the full \texttt{TaylorF2Ecc} model, as well as with Ref.~\cite{RomeroShaw2020}, which discusses possible formation channels for this event. Specifically, Ref.~\cite{RomeroShaw2020} probed GW190425 for residual eccentricity at $10$~Hz to explore channels influenced by unstable case-BB mass transfer. Their bound of $e_0\leq 0.007$ provides no evidence for or against that scenario. Ref.~\cite{LNB} reiterates this point, reporting $e_0\leq 0.023$ at $10$~Hz with a log-uniform prior, and states that this weaker limit does not change the conclusions of Ref.~\cite{RomeroShaw2020}. At $20$~Hz, their analysis yields $e_0 \leq 0.016$, which is in close agreement with our own log-uniform prior result of $e_0 \leq 0.018$. Consequently, we reach the same conclusions as Refs.~\cite{LNB, RomeroShaw2020}. Additionally, our Bayes factor comparison shows no preference for the eccentric model over the quasicircular model, consistent with the expected near-degeneracy of eccentric and quasicircular templates at very low $e_0$.

Our results indicate that GW170817 and GW190425 show no sign of dynamical formation; their negligible $e_0$ values at $20$~Hz are fully consistent with the binaries having formed in the field. Several considerations from earlier work support this view. Typical LVK sources should circularize to $e_0 \lesssim 10^{-4}$ before entering the LIGO--Virgo band~\cite{Peters64, Kowalska11}. Furthermore, only a small fraction (0.2\%--2\%) of BNS are expected to have $e_0 \gtrsim 0.1$ at 30 Hz for Advanced LIGO/Virgo ~\cite{Kowalska11}. Finally, dynamically formed eccentric BNS in globular clusters are expected to be exceedingly rare ($\sim 0.02\,{\rm Gpc}^{-3}\,{\rm yr}^{-1}$)~\cite{Ye2019}. It would nevertheless be valuable to constrain GW170817's $e_0$ at $10$~Hz using an improved \texttt{TaylorF2Ecck} implementation combined with faster analysis techniques (e.g., relative binning~\cite{zackay2018}). In contrast, such lower-frequency analyses will prove more challenging for GW190425 due to its lower network SNR, a limitation already evident in its broader $e_0$ posteriors.

There are many directions to extend the present effort.
It will be interesting to include higher-order $e_0$ contributions in both the Fourier phases and amplitudes of various harmonics. Many of these terms are available in Ref.~\cite{TGHH}, and their data-analysis implications should be explored for ground- and space-based GW observatories~\cite{Paul2023, BBG23}.
Another direction is to develop an efficient template bank for detecting such binaries, as discussed in Ref.~\cite{Phukon2025}.
It will be worthwhile to incorporate spin effects into our approximant. Many required inputs appear in Refs.~\cite{Henry2023, sridhar2024}.
It will also be desirable to add merger and ringdown to our pure inspiral templates by stitching the approximant to numerical-relativity inputs, for example, as reviewed in Ref.~\cite{ficarra2025}.
Detailed parameter-estimation studies relevant to ground- and space-based GW observatories should be pursued for these inspiral models~\cite{Paul2023, BBG23}.
Collectively, however, these efforts call for a \texttt{TaylorF2Ecck} implementation that incorporates $e_0$ through 3.5PN order, which requires extending the 3PN work of Ref.~\cite{Arun09} to 3.5PN accuracy.

During the finalization of this manuscript, the \texttt{pyEFPE} waveform model became available~\cite{Morras2025a}. This is a new \prdsoftware{Python} implementation of the long-developing EFPE framework~\cite{Klein2010, Klein_2018, Klein2021} for modeling inspiraling compact binaries with eccentricity and spin precession, which now also incorporates periastron advance.
This model was recently used for parameter estimation of the GW200105 event (an NSBH binary analyzed as a purely inspiral event by limiting $f_{\rm max}$ to $280$~Hz), claiming evidence of periastron advance and reporting eccentricity $e_0 = 0.145^{+0.007}_{-0.097}$ at $20$~Hz~\cite{Morras2025b}. Although our current study with \texttt{TaylorF2Ecck} focuses on inspiral-dominated BNS events with generally low residual eccentricity ($\lesssim 0.1$), we performed a parameter-estimation run on the same event for comparison. A direct, one-to-one comparison between \texttt{TaylorF2Ecck} and \texttt{pyEFPE} is challenging because the implementations differ in how the gravitational-wave response is obtained.
\texttt{TaylorF2Ecck}, following Ref.~\cite{TGHH}, analytically implements eccentricity-induced orbital effects, including periastron advance and GW emission, and converts these to the frequency domain using the SPA with fully analytic PN terms. In contrast, \texttt{pyEFPE}, as described in Ref.~\cite{Morras2025a}, incorporates eccentricity and periastron advance in a manner that does not allow for a fully analytic frequency-domain eccentric model, owing to numerical integration and other non-analytic approximations in the eccentricity evolution. 
A preliminary mismatch comparison (Appendix~\ref{sec:comparison_pyefpe_taylorf2_family}) shows that \texttt{pyEFPE} diverges significantly from the eccentric \texttt{TaylorF2} family, with ${\cal M}^{*}_{\%} \gtrsim 3$ for $e_0 \gtrsim 0.08$. This is likely due to \texttt{TaylorF2Ecc}/\texttt{TaylorF2Ecck} being limited to leading-order ${\cal O}(e_0^2)$ Fourier phase corrections, whereas \texttt{pyEFPE} incorporates corrections up to ${\cal O}(e_0^4)$, which are important for long inspirals and moderate $e_0$ (see also Ref.~\cite{TGHH} for a comparison of ${\cal O}(e_0^2)$ and ${\cal O}(e_0^6)$ corrections in eccentric inspirals). Future updates to \texttt{TaylorF2Ecck} will enable direct tests of these higher-order corrections.

For GW200105, using the same parameter-estimation settings as in Ref.~\cite{Morras2025b}, our \texttt{TaylorF2Ecck} analysis yielded $e_0 = 0.122^{+0.027}_{-0.098}$ at $20$~Hz, while \texttt{TaylorF2Ecc} gave $e_0 = 0.122^{+0.028}_{-0.096}$. These results are consistent with those reported in Ref.~\cite{Morras2025b} within the 90\% credible intervals. We also ran parameter estimation using \texttt{TaylorF2} approximant with 3PN and 3.5PN accurate phasing. The Bayes-factor comparison gave $\Delta \log_{10} {\cal B} < 1$ when comparing \texttt{TaylorF2Ecck} to both \texttt{TaylorF2Ecc} and \texttt{TaylorF2} for this NSBH event. Consequently, although our eccentricity estimates agree with Ref.~\cite{Morras2025b}, we do not find strong Bayesian evidence for periastron advance or eccentricity using \texttt{TaylorF2Ecck}. Such systems nevertheless remain promising targets for future work, in particular with an improved \texttt{TaylorF2Ecck} implementation.

\section*{Acknowledgments}\label{acknowledgements}

It is a pleasure to thank Otto A. Hannuksela for his suggestions and useful discussions on the parameter-estimation results. We also thank Chunglee Kim for providing further information on the \texttt{TaylorF2Ecc} waveform approximant. The computational work for this manuscript was carried out using resources provided by the LIGO Data Grid (LDG) cluster at the California Institute of Technology (CIT) in California, USA, and the Inter-University Centre for Astronomy and Astrophysics (IUCAA) in Pune, India, also known as Sarathi. We used the \prdsoftware{Python} packages \textsc{bilby\_pipe}~\cite{bilby_pipe_paper} for parameter estimation, \texttt{pycbc}~\cite{pycbc} for match calculations, and \prdsoftware{LALSuite}~\cite{lalsuite} for waveform generation. P.\ H. gratefully acknowledges the support of the Department of Physics at The Chinese University of Hong Kong through the Postgraduate Studentship that made this research possible. P.\ H. also acknowledges support from the Research Grants Council of Hong Kong (Projects No. CUHK 14304622 and No. 14307923), the start-up grant from The Chinese University of Hong Kong, and the Direct Grant for Research from the Research Committee of The Chinese University of Hong Kong. A.\ G. acknowledges support from the Department of Atomic Energy, Government of India, under Project Identification No. RTI 4002, and from the CAS President's International Fellowship Initiative under Grant No. 2026PVA0020. T.\ G.\ F.\ L. is supported by grants from The Research Foundation--Flanders (FWO; Grants No. I002123N and No. I000725N) and the Special Research Fund (BOF; Grant No. STG/21/061). The authors gratefully acknowledge the computational resources and data provided by the LIGO Laboratory and supported by National Science Foundation Grants No. PHY-0757058 and No. PHY-0823459. This research has made use of data or software obtained from the Gravitational Wave Open Science Center (gwosc.org), a service of the LIGO Scientific Collaboration, the Virgo Collaboration, and KAGRA. This material is based upon work supported by NSF's LIGO Laboratory which is a major facility fully funded by the National Science Foundation, as well as the Science and Technology Facilities Council (STFC) of the United Kingdom, the Max-Planck-Society (MPS), and the State of Niedersachsen/Germany for support of the construction of Advanced LIGO and construction and operation of the GEO600 detector. Additional support for Advanced LIGO was provided by the Australian Research Council. Virgo is funded, through the European Gravitational Observatory (EGO), by the French Centre National de Recherche Scientifique (CNRS), the Italian Istituto Nazionale di Fisica Nucleare (INFN) and the Dutch Nikhef, with contributions by institutions from Belgium, Germany, Greece, Hungary, Ireland, Japan, Monaco, Poland, Portugal, Spain. K.\ A.\ G.\ R.\ A. is supported by Ministry of Education, Culture, Sports, Science and Technology (MEXT), Japan Society for the Promotion of Science (JSPS) in Japan; National Research Foundation (NRF) and Ministry of Science and ICT (MSIT) in Korea; Academia Sinica (AS) and National Science and Technology Council (NSTC) in Taiwan.

\section*{Data availability}\label{DataAvailability}

This article's data, code, and plot generation Jupyter-lab-notebook are accessible in the associated data release on GitHub~\cite{TaylorF2EcckRepo}. The version of \prdsoftware{LALSuite} that includes \texttt{TaylorF2Ecck}, \texttt{TaylorF2Ecch}, and the extended PN-order \texttt{TaylorF2} is available in a forked GitLab repository~\cite{lalsuitehemanta2022} from the original repository~\cite{lalsuite}.

For both GW170817 and GW190425, we use publicly available gravitational-wave strain data from the Advanced LIGO and Virgo detectors, obtained through the LIGO Open Science Center (LOSC)~\cite{Vallisneri_2015}. To ensure consistency with the original LVK parameter-estimation analyses~\cite{GW170817pe, Abbott2020GW190425}, we use the official detector-specific \texttt{BayesWave}~\cite{bayeswave} power spectral densities released with the corresponding data products~\cite{LIGO-P1800061, LIGO2019_GW190425_data, LIGO-P2000026}.

For GW170817, we analyze the \texttt{LOSC\_CLN\_4\_V1} data set for the H1, L1, and V1 detectors~\cite{Blackburn2017_LOSC_CLN, Driggers2019}. The frame channels are
\texttt{\{H1:LOSC-STRAIN, L1:LOSC-STRAIN, V1:LOSC-STRAIN\}}.
This data set includes the noise-subtracted strain released by the LVK Collaboration. Additional details of this release are given in Ref.~\cite{GWOSC2017_GW170817}.

For GW190425, we use the \texttt{T1700406\_v3} data set for the L1 and V1 detectors~\cite{Abbott2020b_GW190425Glitch}. The frame channels are
\texttt{\{L1:DCS-CALIB\_STRAIN\_CLEAN\_C01\_T1700406\_v3, V1:Hrec\_hoft\_16384Hz\_T1700406\_v3\}}.
This data set includes preprocessing glitch removal performed by the LVK Collaboration for parameter estimation. Further details are provided in the GW190425 data release and parameter-estimation sample release in Refs.~\cite{LIGO2019_GW190425_data, LIGO-P2000026, Abbott2020GW190425}.\\

\appendix

\renewcommand{\theequation}{A\arabic{equation}} 

\section{Detailed analytical structure of the \texttt{TaylorF2Ecck} waveform approximant}\label{sec:TaylorF2Ecck_full}

The following subsections provide the analytical structure of the \texttt{TaylorF2Ecck} waveform approximant. We present the frequency-domain waveform $\tilde{h}(f)$, the Fourier phase $\Psi_{j,n}$ for each harmonic $(j,n)$, the periastron-advance parameter $k$, and the eccentricity evolution $e_t$, following inputs from Ref.~\cite{TGHH}.\\

\subsection{Analytical expression for the frequency-domain waveform $\tilde{h}(f)$ at 3PN accuracy}\label{sec:TaylorF2Ecck_hf}

Below, we present the explicit expression for $\tilde{h}(f)$ with a 3PN-accurate Fourier phase $\Psi_{j,n}$, including leading-order ${\cal O}(e_0^2)$ eccentric corrections as given in Eq.~(\ref{eq:Ecck_psi_full}), and a quadrupolar-order amplitude with leading-order ${\cal O}(e_0)$ eccentric corrections as given in Eq.~(\ref{eq:xiEcck_simple}). The expression reads
\begin{subequations}
\begin{align}
\tilde{h}(f) = & \, \left( \frac{5\pi \eta}{384} \right)^{1/2} \frac{(G m)^2}{c^5 d_L} \left(\frac{Gm\pi f}{c^3}\right)^{-7/6} \notag \\*[2.0bp]
&\times \sum_{j,n} \xi_{j,n} \left(\frac{j}{2}\right)^{2/3} e^{-i(\Psi_{j,n}-\pi/4)} \Theta_{j,n}\,,
\end{align}
\noindent with 
\begin{align}
\Theta_{j,n} = & \, \text{Heaviside}\Bigl[ \left(j-(j+n)\frac{k}{1+k} \right)f_{\text{LSO}} - 2f \Bigr]\,, \notag \\*[2.0bp]
=&\, 
    \begin{cases}
        1 & \text{if } \left(j-(j+n)\frac{k}{1+k} \right)f_{\text{LSO}} > 2f\,,\\
        0 & \text{if } \left(j-(j+n)\frac{k}{1+k} \right)f_{\text{LSO}} \leq 2f\,,
    \end{cases}
\end{align}
\end{subequations}
\noindent where $f_{\rm LSO}$ is the orbital frequency at the last stable orbit, given by
\onecolumngrid
\begin{align}
    f_{\rm LSO} = &\, \frac{c^3}{G m \pi 6^{3/2}}\,. \label{eq:flso}
\end{align}\\

\subsection{Explicit 3PN-order expression for the\\ Fourier phase $\Psi_{j,n}$}\label{sec:TaylorF2Ecck_psi}

Here we present the analytical form of the Fourier phase $\Psi_{j,n}$ for each $(j,n)$ harmonic. It is accurate to 3PN order and includes orbital-eccentricity contributions up to ${\cal O}(e_0^2)$. The expression reads

\begin{subequations}
\label{eq:Ecck_psi_full}
\begin{align}
&\Psi_{j,n} = \, -2 \pi  f t_c+\phi_c \left(j-(j+n)\frac{k}{1+k} \right)
 - \frac{3 j}{256 \eta  x_{j,n}^{5/2}} \sum _{u=0}^6 {\cal D}_u\,x_{j,n}^{u/2}\,,\label{eq:Ecck_psi_full_}
\end{align}

\noindent with
\begin{align}
&x_{j,n} = \, \left[ \frac{2 \pi G m f}{\left\{j - \frac{(j+n)k}{1+k}\right\}c^3} \right]^{2/3}\,. 
\end{align}
\end{subequations}
The PN coefficients ${\cal D}_u$ in Eq.~(\ref{eq:Ecck_psi_full}) are

\begin{subequations}
\begin{align}
&{\cal D}_0= \, 1- e_0^2 \left\{ \frac{2355}{1462} \chi^{-19/9}\right\}\,, \\[20.0bp]
&{\cal D}_1= \, 0\,, \\[20.0bp]
&{\cal D}_2= \, \frac{55 \eta }{9}-\frac{25 n}{3 j}-\frac{2585}{756} + e_0^2 \left\{\left(-\frac{128365 \eta }{12432}+\frac{1805 n}{172 j}+\frac{69114725}{14968128}\right) \chi^{-19/9} + \left(\frac{154645 \eta }{17544}-\frac{2223905}{491232}\right) \chi^{-25/9} \right\}\,, \\[20.0bp]
&{\cal D}_3= \, -16 \pi + e_0^2 \Bigg\{\frac{65561 \pi }{4080}\chi ^{-19/9}-\frac{295945 \pi }{35088}\chi ^{-28/9}\Bigg\}\,, \\[20.0bp]
&{\cal D}_4= \, \frac{3085 \eta ^2}{72}+\eta  \left(\frac{22105}{504}-\frac{10 n}{j}\right)-\frac{31805 n}{252 j}-\frac{48825515}{508032} \notag \\*[2.0bp]
&\qquad  +e_0^2 \left\{\left(-\frac{10688155 \eta ^2}{294624} + \eta  \left(\frac{36539875 n}{1260072 j}-\frac{72324815665}{6562454976}\right)+\frac{323580365 n}{5040288 j}+\frac{115250777195}{2045440512}\right)\chi ^{-19/9} \right. \notag \\*[2.0bp]
&\qquad \left. + \left(\frac{25287905 \eta ^2}{447552}+\eta  \left(-\frac{355585 n}{6192 j}-\frac{3656612095}{67356576}\right)+\frac{5113565 n}{173376 j}+\frac{195802015925}{15087873024}\right)\chi ^{-25/9} \right. \notag \\*[2.0bp]
&\qquad \left. + \left(-\frac{14251675 \eta ^2}{631584}+\frac{3062285 \eta }{260064}+\frac{936702035}{1485485568}\right)\chi ^{-31/9} \right\}\,, \displaybreak[4]\\[20.0bp]
 &{\cal D}_5= \, -\left(\frac{65 \pi \eta }{9}+\frac{160 \pi n}{3 j}+\frac{1675 \pi }{756}\right) \left(\log \left(\frac{f}{f_\text{LSO}}\right)-\log (j)\right)-\frac{1}{9} (65 \pi ) \eta -\frac{32 \pi n}{j}+\frac{14453 \pi }{756} \notag \\*[2.0bp]
 &\qquad  + e_0^2 \left\{ \left(\frac{15803101 \pi \eta }{229824}-\frac{4909969 \pi n}{46512 j}-\frac{458370775 \pi }{6837264}\right) \chi^{-19/9} \right. \notag \\*[2.0bp]
 &\qquad  + \left(-\frac{48393605 \pi \eta }{895104}+\frac{680485 \pi n}{12384 j}+\frac{26056251325 \pi }{1077705216}\right) \chi^{-28/9} \notag \\*[2.0bp]
 &\qquad  \left. + \left(\frac{185734313 \pi }{4112640}-\frac{12915517 \pi \eta }{146880}\right) \chi^{-25/9} + \left(\frac{149064749 \pi \eta }{2210544}-\frac{7063901 \pi }{520128}\right) \chi^{-34/9} \right\}\,,  \\[25.0bp]
&{\cal D}_6= -\frac{6848 \gamma }{21}-\frac{127825 \eta ^3}{1296}+\eta ^2 \left(\frac{475 n}{24 j}+\frac{110255}{1728}\right) \notag \\*[2.0bp]
&\qquad  + \eta  \left(\frac{1845 \pi ^2 n}{32 j}-\frac{2393105 n}{1512 j}+\frac{23575 \pi ^2}{96}-\frac{20562265315}{3048192}\right)+\frac{257982425 n}{508032 j}-\frac{3424 \log (x_{j,n})}{21}-\frac{640 \pi ^2}{3} \notag \\*[2.0bp]
&\qquad  +\frac{13966988843531}{4694215680}-\frac{13696 \log (2)}{21} + e_0^2 \left\{ \left(\frac{2105566535 \eta ^3}{10606464}+\eta ^2 \left(-\frac{7198355375 n}{45362592 j}-\frac{2186530635995}{52499639808}\right) \right. \right. \notag \\*[2.0bp]
&\qquad \left. +\eta  \left(-\frac{9519440485 n}{35282016 j}-\frac{13467050491570355}{39689727694848}\right)+\frac{916703174045 n}{5080610304 j}+\frac{326505451793435}{2061804036096}\right) \chi^{-25/9} \notag \\*[2.0bp]
&\qquad  + \left(-\frac{2330466575 \eta ^3}{16111872}+\eta ^2 \left(\frac{32769775 n}{222912 j}+\frac{906325428545}{6466231296}\right) \right. \notag \\*[2.0bp]
&\qquad \left. +\eta  \left(-\frac{119702185 n}{1560384 j}-\frac{48415393035455}{1629490286592}\right)-\frac{2153818055 n}{524289024 j}-\frac{82471214720975}{45625728024576}\right) \chi^{-31/9} \notag \\*[2.0bp]
&\qquad  + \left(-\frac{734341 \gamma }{16800}-\frac{69237581 \eta ^3}{746496}+\eta ^2 \left(\frac{43766986495 n}{1022720256 j}-\frac{159596464273381}{1718170030080}\right) \right. \notag \\*[2.0bp]
&\qquad  +\eta  \left(\frac{639805 \pi ^2 n}{22016 j}-\frac{1219797059185 n}{2045440512 j}+\frac{12111605 \pi ^2}{264192}-\frac{37399145056383727}{28865256505344}\right)+\frac{534109712725265 n}{2405438042112 j} \notag \\*[2.0bp]
&\qquad \left. -\frac{734341 \log (x_{j,n})}{33600}-\frac{21508213 \pi ^2}{276480}+\frac{4175723876720788380517}{5556561877278720000} +\frac{4602177 \log (3)}{44800}-\frac{9663919 \log (2)}{50400}\right) \chi^{-19/9} \notag \\*[2.0bp]
&\qquad  +\frac{24716497 \pi ^2}{293760}\chi^{-28/9} + \left(\frac{2603845 \gamma }{61404}+\frac{2425890995 \eta ^3}{68211072}+\frac{4499991305 \eta ^2}{636636672} \right. \notag \\*[2.0bp]
&\qquad  +\left(\frac{3121945 \pi ^2}{561408}-\frac{1437364085977}{53477480448}\right) \eta -\frac{2603845 \log (\chi )}{184212}+\frac{2603845 \log (x_{j,n})}{122808} -\frac{96423905 \pi ^2}{5052672} \notag \\*[2.0bp]
&\qquad \left. \left. -\frac{4165508390854487}{16471063977984}+\frac{1898287 \log (2)}{184212}+\frac{12246471 \log (3)}{163744}\right) \chi^{-37/9} \right\}\,.
\end{align}
\end{subequations}
\noindent\hfill\prdwideclosingrule\par\vspace{8bp}
\twocolumngrid

We also present the \prdsoftware{LALSuite}-compatible form of $\Psi_{j,n}$, where the reference phase $\phi_{\rm ref}$, defined at a chosen reference frequency $f_{\rm ref}$, is used instead of $\phi_c$. Defining the last term in Eq.~(\ref{eq:Ecck_psi_full_}) as
\begin{align}
\psi_{j,n} = \frac{3 j}{256 \eta  x_{j,n}^{5/2}} \sum _{u=0}^6 {\cal D}_u\,x_{j,n}^{u/2} \,,
\end{align}
\noindent and requiring $\Psi_{j,n}$ to equal $\Psi_{j,n}^{\rm ref}$ at $f_{\rm ref}$, the phase-shifted form reads

\begin{align}
\label{e:lal_psi_jn}
\Psi_{j,n} =& \, -2 \pi  f t_c+\phi_{\rm ref} \left(j-(j+n)\frac{k}{1+k} \right) + \psi_{j,n}^{\rm ref} - \psi_{j,n}\,.
\end{align}

\subsection{Explicit 3PN-order expression for the periastron advance parameter $k$}\label{sec:periastron_k}

The complete expression for the periastron advance parameter $k$ is given below. It includes terms up to 3PN order and leading-order eccentricity corrections ${\cal O}(e_0^2)$. The
\noindent expression for $k$ reads
\begin{subequations}
\label{eq:k_full}
\begin{align}
&k_j = \, \sum_{u=0}^6 {\cal K}_u \, x^{u/2}_j \,,
\end{align}
\noindent with
\begin{align}
&x_j = \, \left(\frac{2 \pi G m f}{j c^3}\right)^{2/3}\,. 
\end{align}
\end{subequations}
The coefficients in Eq.~(\ref{eq:k_full}) are listed below.

\begin{subequations}
\label{eq:k_coeff}
\begin{align}
&{\cal K}_{0} = \, 0\,, \\[10.0bp]
&{\cal K}_{1} = \, 0\,, \\[10.0bp]
&{\cal K}_{2} = \, 3 + e_0^2 \left\{ \frac{3}{\chi^{19/9}} \right\}\,, \\[10.0bp]
&{\cal K}_{3} = \, 0 \,,
\end{align}
\onecolumngrid
\par\nointerlineskip\hbox{\prdwideopeningrule}
\begin{align}
&{\cal K}_{4} = \, \frac{27}{2} - 7\eta + e_0^2 \left\{ \left(\frac{2833}{336} - \frac{197\eta}{12}\right)\frac{1}{\chi^{25/9}} + \left(\frac{10523}{336} - \frac{49\eta}{12}\right)\frac{1}{\chi^{19/9}} \right\}\,, \\[10.0bp]
&{\cal K}_{5} = \, e_0^2 \left\{ \frac{1}{\chi^{28/9}} + \frac{1}{\chi^{19/9}} \right\}\frac{377\pi}{24}\,, \\[10.0bp]
&{\cal K}_{6} = \, \frac{1}{32}(2160 +( 123\pi^2 - 5192)\eta + 224\eta^2) + e_0^2 \left\{ \left(-\frac{1193251}{1016064} -\frac{66317\eta}{3024} +\frac{18155\eta^2}{432}\right)\frac{1}{\chi^{31/9}} \right. \notag \\*[2.0bp] 
&\qquad \left. + \left(\frac{29811659}{338688} -\frac{276481\eta}{1512} +\frac{9653\eta^2}{432}\right)\frac{1}{\chi^{25/9}} + \left(\frac{142961921}{508032} -\frac{1419653\eta }{3024} +\frac{1599 \pi^2\eta}{128} +\frac{91\eta^2}{27} \right)\frac{1}{\chi^{19/9}} \right\}\,.
\end{align}
\end{subequations}

\subsection{Explicit expression for eccentricity evolution $e_t$}\label{sec:et_evolution}

Following Appendix C of Ref.~\cite{TGHH}, our 3PN-${\cal O}(e_0)$ accurate expression for $e_t$ as a function of frequency reads
\begin{align}
e_t =& \, \sum_{m=0}^{6} {\cal E}_m \, x_{j,n}^{m/2},\label{eq:et}
\end{align}
\noindent with 
\begin{subequations}
\begin{align}
&{\cal E}_0= \,e_0\, \chi^{-19/18}\,, \\[10.0bp]
&{\cal E}_1= \,0\,, \\[10.0bp]
&{\cal E}_2=\, e_0 \left\{\left(-\frac{2833}{2016}+\frac{197 \eta }{72}\right)\chi^{-19/18}+\left(\frac{2833}{2016}-\frac{197\eta}{72}\right)\chi^{-31/18}\right\}\,, \\[10.0bp]
&{\cal E}_3= \, e_0\left\{\frac{377}{144}\pi\left(-\chi^{-19/18}+\chi ^{-37/18}\right)\right\}\,, \\[10.0bp]
&{\cal E}_4= \,e_0 \left\{\left(\frac{77006005}{24385536}-\frac{1143767 \eta }{145152}+\frac{43807 \eta ^2}{10368}\right)\chi^{-19/18}+\left(-\frac{8025889}{4064256}+\frac{558101\eta }{72576}-\frac{38809 \eta^2}{5184}\right)\chi^{-31/18} \right. \notag \\*[2.0bp] 
&\qquad \left. +\left(-\frac{28850671}{24385536}+\frac{27565 \eta }{145152}+\frac{33811 \eta^2}{10368}\right) \chi ^{-43/18}\right\}\,, \\[10.0bp]
&{\cal E}_5= \, e_0 \left\{\left(\frac{9901567 \pi }{1451520}-\frac{202589 \pi  \eta }{362880}\right) \chi ^{-19/18}+\left(-\frac{1068041 \pi }{290304}+\frac{74269 \pi  \eta}{10368}\right) \chi ^{-31/18} \right. \notag \\*[2.0bp] 
&\qquad  \left. +\left(-\frac{1068041 \pi }{290304}+\frac{74269 \pi  \eta }{10368}\right) \chi ^{-37/18} +\left(\frac{778843 \pi}{1451520}-\frac{4996241 \pi  \eta }{362880}\right) \chi ^{-49/18}\right\}\,, \\[10.0bp]
&{\cal E}_6= \, \left\{\left(-\frac{33320661414619}{386266890240}+\frac{180721 \pi^2}{41472}+\frac{3317 \gamma }{252}+\left(\frac{161339510737}{8778792960}+\frac{3977 \pi ^2}{2304}\right) \eta -\frac{359037739 \eta^2}{20901888} +\frac{10647791 \eta ^3}{2239488} \right. \right. \notag \\*[2.0bp]
&\qquad \left. +\frac{12091 \log (2)}{3780}+\frac{26001 \log (3)}{1120}+\frac{3317 \log (x_{j,n})}{504}\right)\chi ^{-19/18} +\left(\frac{218158012165}{49161240576}-\frac{34611934451 \eta }{1755758592}+\frac{191583143 \eta ^2}{6967296} \right. \notag \\*[2.0bp]
&\qquad \left. -\frac{8629979 \eta ^3}{746496}\right) \chi^{-31/18}-\frac{142129 \pi ^2}{20736}\chi ^{-37/18} +\left(\frac{81733950943}{49161240576}-\frac{6152132057 \eta }{1755758592}-\frac{1348031 \eta^2}{331776} \right. \notag \\*[2.0bp]
&\qquad \left. +\frac{6660767 \eta ^3}{746496}\right) \chi ^{-43/18}+\left(\frac{216750571931393}{2703868231680}+\frac{103537 \pi ^2}{41472}-\frac{3317 \gamma }{252} +\left(\frac{866955547}{179159040}-\frac{3977 \pi^2}{2304}\right) \eta \right.\notag \\*[2.0bp] 
&\qquad \left.\left. -\frac{130785737 \eta ^2}{20901888}-\frac{4740155 \eta ^3}{2239488}-\frac{12091 \log (2)}{3780}-\frac{26001 \log(3)}{1120} -\frac{3317 \log (x_{j,n})}{504}-\frac{3317 \log (\chi )}{756}\right)\chi ^{-55/18}\right\}e_0 \,.
\end{align}
\end{subequations}
\noindent\hfill\prdwideclosingrule\par\vspace{2bp}
\renewcommand{\theequation}{B\arabic{equation}} 
\begingroup
\setlength{\prddisplayskip}{4bp}
\makeatletter\prd@setdisplayskips\makeatother
\patchcmd{\section}{15bp plus4bp minus4bp}{12bp}{}%
  {\PackageError{prd-layout}{Appendix B section-spacing patch failed}{Check the section definition.}}
\patchcmd{\subsection}{13bp plus4bp minus4bp}{12bp}{}%
  {\PackageError{prd-layout}{Appendix B heading-spacing patch failed}{Check the subsection definition.}}
\noindent\begin{minipage}[t]{\dimexpr(\textwidth-\columnsep)/2\relax}
\vspace{0pt}
\section{Detailed analytical structure of the \texttt{TaylorF2Ecch} waveform approximant}\label{sec:TaylorF2Ecch_full}

The following subsections provide the analytical structure of the \texttt{TaylorF2Ecch} waveform approximant. We present the frequency-domain waveform $\tilde{h}(f)$ and the associated Fourier phase $\Psi_j$.

\subsection{Analytical expression for the frequency-domain waveform $\tilde{h}(f)$ at 3PN accuracy}\label{sec:TaylorF2Ecch_hf}

Below, we present the explicit expression for $\tilde{h}(f)$ with a 3PN-accurate Fourier phase $\Psi_{j}$, including leading-order ${\cal O}(e_0^2)$ eccentric corrections as given in Eq.~(\ref{eq:Ecch_psi_full}), and a quadrupolar-order amplitude with leading-order ${\cal O}(e_0)$ eccentric corrections as given in Eq.~(\ref{eq:xiEcch_simple}).
\noindent The expression reads

\end{minipage}\hfill
\begin{minipage}[t]{\dimexpr(\textwidth-\columnsep)/2\relax}
\vspace{0pt}
\begin{align}
\tilde{h}(f) = & \, \left( \frac{5\pi \eta}{384} \right)^{1/2} \frac{(G m)^2}{c^5 d_L} \left(\frac{Gm\pi f}{c^3}\right)^{-7/6} \notag \\*[2.0bp]
&\times \sum_{j} \xi_{j} \left(\frac{j}{2}\right)^{2/3} e^{-i(\Psi_{j}-\pi/4)} \Theta_{j}\,,
\end{align}
\noindent where $\Theta_{j} = \text{Heaviside}\left[j f_{\text{LSO}}-2f \right]$.

\subsection{Explicit 3PN-order expression for the Fourier phase $\Psi_{j}$}\label{sec:TaylorF2Ecch_psi}

The analytical form of the 3PN-order Fourier phase $\Psi_{j}$ for each $j$th harmonic, including ${\cal O}(e_0^2)$ contributions, is given by
\begin{align}
\label{eq:Ecch_psi_full}
\Psi_{j} =& \, -2 \pi  f t_c+\phi _c j - \frac{3 j}{256 \eta  x_{j}^{5/2}} \sum _{u=0}^6 {\cal D}_u x_j^{u/2}\,,
\end{align}
\end{minipage}\par
\endgroup
\vspace{20bp}\nointerlineskip\hbox{\prdwideopeningrule}
\vspace{20bp}
\noindent where $x_{j} = \left( \frac{2 \pi G m f}{jc^3} \right)^{2/3}$. The PN coefficients ${\cal D}_u$ in Eq.~(\ref{eq:Ecch_psi_full}) are
\begin{subequations}
\begin{align}
&{\cal D}_0= \, 1- e_0^2 \left\{ \frac{2355}{1462} \chi^{-19/9}\right\}\,, \\[10.0bp]
&{\cal D}_1= \, 0\,, \\[10.0bp]
&{\cal D}_2= \, \frac{3715}{756} + \frac{55 \eta}{9} + e_0^2 \left\{ \left( -\frac{2223905}{491232} + \frac{154645 \eta}{17544} \right) \chi^{-25/9} + \left( -\frac{2045665}{348096} - \frac{128365 \eta}{12432} \right) \chi^{-19/9} \right\} \,, \\[10.0bp] 
&{\cal D}_3= \, -16 \pi + e_0^2 \left\{ -\frac{295945 \pi}{35088} \chi^{-28/9} + \frac{65561 \pi}{4080} \chi^{-19/9} \right\} \,, \\[10.0bp]
&{\cal D}_4= \, \frac{15293365}{508032} + \frac{27145 \eta}{504} + \frac{3085 \eta^2}{72} + e_0^2 \left\{ \left( \frac{936702035}{1485485568} + \frac{3062285 \eta}{260064} - \frac{14251675 \eta^2}{631584} \right) \chi^{-31/9} \right. \notag\\*[2.0bp]
&\qquad \left. + \left( -\frac{5795368945}{350880768} + \frac{4917245 \eta}{1566432} + \frac{25287905 \eta^2}{447552} \right) \chi^{-25/9} + \left( -\frac{111064865}{14141952} - \frac{165068815 \eta}{4124736} - \frac{10688155 \eta^2}{294624} \right) \chi^{-19/9} \right\} \,, \\[10.0bp]
&{\cal D}_5= \, \left( \frac{38645}{756} - \frac{65 \eta}{9} \right) \pi \left( 1 + \log\left(x_j^{3/2}\right) \right) + e_0^2 \left\{ \left( -\frac{7063901 \pi}{520128} + \frac{149064749 \pi \eta}{2210544} \right) \chi^{-34/9} \right. \notag \\*[2.0bp]
&\qquad  + \left( -\frac{771215705 \pi}{25062912} - \frac{48393605 \pi \eta}{895104} \right) \chi^{-28/9} + \left( \frac{185734313 \pi}{4112640} - \frac{12915517 \pi \eta}{146880} \right) \chi^{-25/9} \notag \\*[2.0bp]
&\qquad \left. + \left( \frac{3873451 \pi}{100548} + \frac{15803101 \pi \eta}{229824} \right) \chi^{-19/9} \right\} \,,\\[10.0bp]
&{\cal D}_6= \, \frac{11583231236531}{4694215680} - \frac{640 \pi^2}{3} - \frac{6848 \gamma}{21} + \left( -\frac{15737765635}{3048192} + \frac{2255 \pi^2}{12} \right) \eta + \frac{76055 \eta^2}{1728} - \frac{127825 \eta^3}{1296} - \frac{13696 \log(2)}{21} \notag\\*[2.0bp]
&\qquad - \frac{3424 \log(x_j)}{21} + e_0^2 \left\{ \left( \frac{2440991806915}{1061063442432} + \frac{1781120054275 \eta}{37895122944} - \frac{1029307085 \eta^2}{150377472} - \frac{2330466575 \eta^3}{16111872} \right) \chi^{-31/9} \right. \notag \\*[2.0bp]
&\qquad  + \frac{24716497 \pi^2}{293760} \chi^{-28/9} + \left( -\frac{314646762545}{14255087616} - \frac{1733730575525 \eta}{24946403328} + \frac{11585856665 \eta^2}{98993664} + \frac{2105566535 \eta^3}{10606464} \right) \chi^{-25/9} \notag \\*[2.0bp]
&\qquad  + \left( \frac{59648637301056877}{112661176320000} - \frac{21508213 \pi^2}{276480} - \frac{734341 \gamma}{16800} + \left( -\frac{409265200567}{585252864} + \frac{103115 \pi^2}{6144} \right) \eta - \frac{4726688461 \eta^2}{34836480} \right. \notag \\*[2.0bp]
&\qquad  \left. - \frac{69237581 \eta^3}{746496} - \frac{9663919 \log(2)}{50400} + \frac{4602177 \log(3)}{44800} - \frac{734341 \log(x_j)}{33600} \right) \chi^{-19/9} \notag \\*[2.0bp]
&\qquad  + \left( -\frac{4165508390854487}{16471063977984} - \frac{96423905 \pi^2}{5052672} + \frac{2603845 \gamma}{61404} + \left( -\frac{1437364085977}{53477480448} + \frac{3121945 \pi^2}{561408} \right) \eta + \frac{4499991305 \eta^2}{636636672} \right. \notag \\*[2.0bp]
&\qquad  \left. \left. + \frac{2425890995 \eta^3}{68211072} + \frac{1898287 \log(2)}{184212} + \frac{12246471 \log(3)}{163744} + \frac{2603845 \log(x_j)}{122808} - \frac{2603845 \log(\chi)}{184212} \right) \chi^{-37/9} \right\}\,.
\end{align}
\end{subequations}
\par\nointerlineskip\hbox to\textwidth{\hfil\prdwideclosingrule}
\vspace{8bp}\twocolumngrid

Similar to Eq.~(\ref{e:lal_psi_jn}), the \prdsoftware{LALSuite}-compatible form of $\Psi_j$ can be written as
\begin{align}
\label{e:lal_psi_j}
\Psi_{j} =& \, -2 \pi  f t_c+\phi_{\rm ref}j + \psi_{j}^{\rm ref} - \psi_{j}\,.
\end{align}

\section{Supplementary sanity checks}
\label{sec:sanity_tests_appendix}

This appendix presents supplementary sanity checks for the waveform comparisons discussed in Sec.~\ref{sec:sanity_check}. We first compare the real part of the plus polarization, $\Re[\tilde{h}_+(f)]$, for the waveform approximants used in this work. We then provide extended match-test results in the eccentricity--mass parameter space, complementing the corresponding discussion in the main text.

\subsection{Waveform comparison using the real part of $\tilde{h}_+(f)$}
\label{sec:real_part_waveform_comparison}

As a supplementary waveform-level sanity check, we compare the frequency evolution of the real part of the plus polarization, $\Re[\tilde{h}_+(f)]$, for the eccentric approximants \texttt{TaylorF2Ecck}, \texttt{TaylorF2Ecch}, and \texttt{TaylorF2Ecc}, together with the quasicircular \texttt{TaylorF2} model. The comparison is performed using simulated GW150914-like BBH parameters with the same setup as in Sec.~\ref{sec:sanity_check}, with an initial eccentricity $e_0=0.1$ at $20~\mathrm{Hz}$. This test complements the $|\tilde{h}_+(f)|$ comparison shown in Fig.~\ref{fig:sanity-check1} by highlighting the oscillatory structure of the frequency-domain waveform. This is useful because the real part of each harmonic contains an oscillatory contribution from the phase factor. For example, in \texttt{TaylorF2Ecck}, this contribution scales schematically as $\Re[e^{-i(\Psi_{j,n}-\pi/4)}]=\cos(\Psi_{j,n}-\pi/4)$. Consequently, differences in the Fourier phases, including those arising from eccentric corrections, appear as frequency-dependent phase shifts in $\Re[\tilde{h}_+(f)]$.

\begin{figure}[t!]
\centering
\centering
    \includegraphics{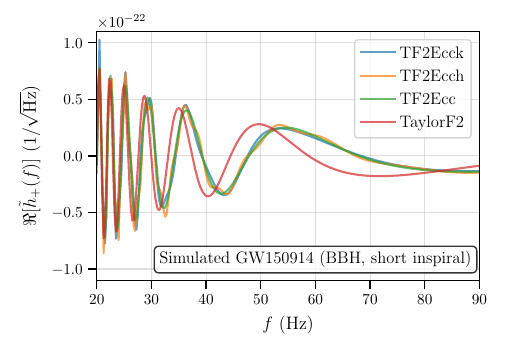}
\caption{Frequency evolution of the real part of the plus polarization, $\Re[\tilde{h}_+(f)]$, for the eccentric \texttt{TaylorF2Ecck}, \texttt{TaylorF2Ecch}, and \texttt{TaylorF2Ecc} approximants, compared with the quasicircular \texttt{TaylorF2} model. The waveforms are generated using simulated GW150914-like BBH parameters with $e_0=0.1$ at $20~\mathrm{Hz}$. For this moderate eccentricity, the eccentric models show broadly similar phase evolution among themselves, while exhibiting clear frequency-dependent phase shifts relative to the quasicircular waveform. The amplitude modulations in \texttt{TaylorF2Ecck} and \texttt{TaylorF2Ecch} are better visualized in the log-scale comparison of Fig.~\ref{fig:sanity-check1}.}\label{fig:real_waveform_gw150914}
\end{figure}

Figure~\ref{fig:real_waveform_gw150914} shows that, for $e_0=0.1$, the three eccentric approximants have broadly similar frequency-domain phase evolution in $\Re[\tilde{h}_+(f)]$ over the plotted range. This similarity is mainly driven by the comparable Fourier-phase evolution of the dominant $(2,-2)$ harmonic across the eccentric models. However, all three eccentric waveforms exhibit visible frequency-dependent phase shifts relative to the quasicircular \texttt{TaylorF2} waveform. This behavior is consistent with the presence of eccentric corrections in the Fourier phases. Distinct amplitude modulations are also visible in this comparison, but they are better shown in the log-scale amplitude plot of Fig.~\ref{fig:sanity-check1}.

\begin{figure}[t!]
    \centering
    \includegraphics{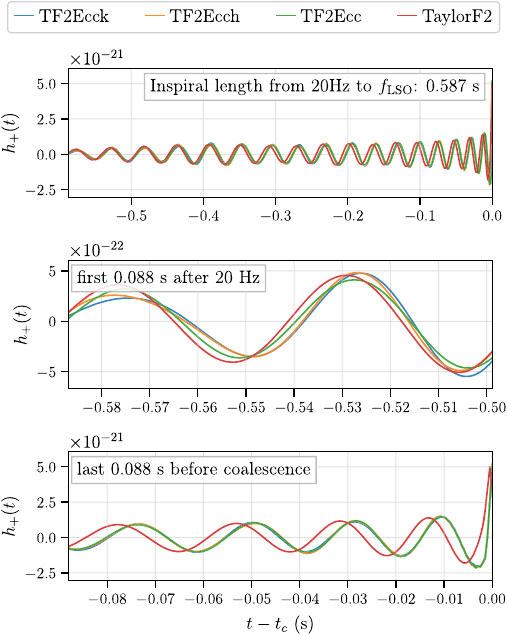}
    \caption{Time-domain evolution of the plus polarization, $h_+(t)$, obtained through an inverse Fourier transform of the waveforms $h_+(f)$ (see Fig.~\ref{fig:real_waveform_gw150914} for $\Re[\tilde{h}_+(f)]$ evolution). Consistent with the frequency-domain comparison, the eccentric approximants show broadly similar time evolution among themselves for this moderate eccentricity. Small amplitude differences are visible at early times but diminish at late times as the eccentricity decreases. The eccentric waveforms also show visible time-dependent phase shifts relative to the quasicircular \texttt{TaylorF2} waveform.}\label{fig:time_waveform_gw150914}
\end{figure}

To provide a complementary visualization, we also show the corresponding time-domain evolution obtained through an inverse Fourier transform of $\tilde{h}_+(f)$. As illustrated in Fig.~\ref{fig:time_waveform_gw150914}, the time-domain representation, $h_+(t)$, is consistent with the frequency-domain comparison. The eccentric approximants show broadly similar time evolution among themselves, with small amplitude differences at early times that diminish as the binary circularizes and the eccentricity decreases. The time-domain representation also shows visible phase shifts between the eccentric waveforms and the quasicircular \texttt{TaylorF2} waveform over the plotted interval.

\subsection{Match tests in the parameter space of eccentricity and mass}\label{sec:match_test_appendix}


\begin{figure*}[ht!]
\centering
\includegraphics{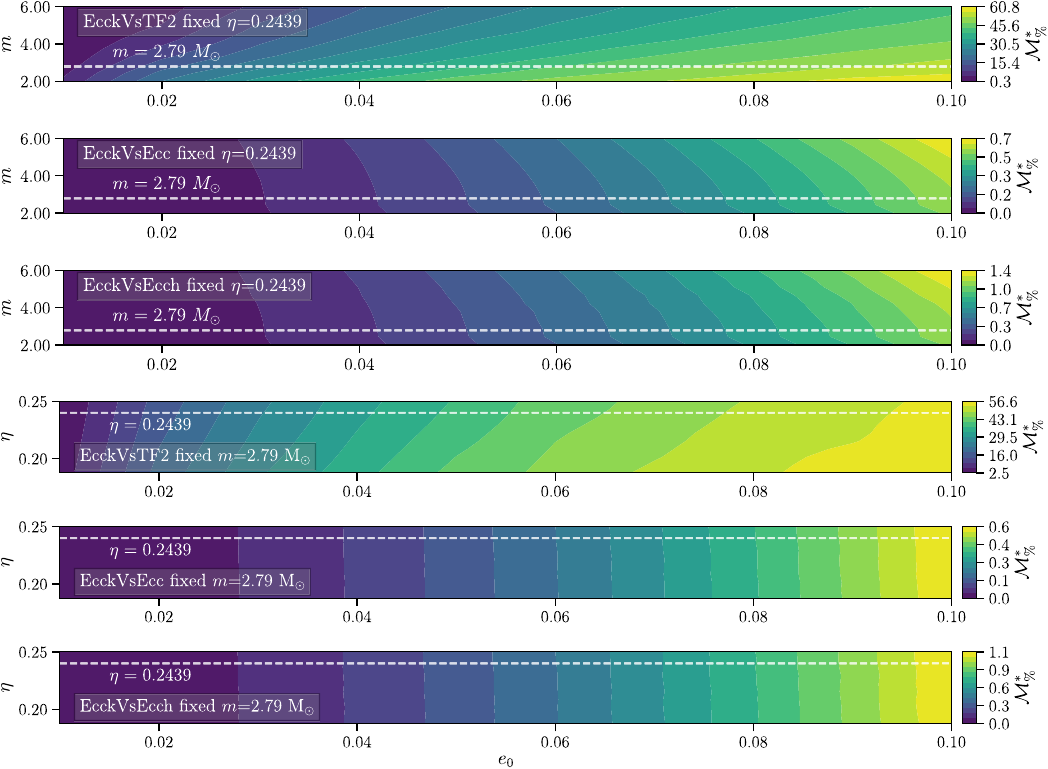}
\caption{Extended waveform similarity analysis complementing Fig.~\ref{fig:contour_mismatch}. The mismatch percentage (${\cal M}^{*}_{\%}$) illustrates waveform deviation, comparing \texttt{TaylorF2Ecck} with eccentric models \texttt{TaylorF2Ecc} and \texttt{TaylorF2Ecch}, and the quasicircular \texttt{TaylorF2} model, across a parameter space relevant for BNS systems. White dashed lines highlight GW170817-like parameters. The top three panels show ${\cal M}^{*}_{\%}$ as a function of $m$ and $e_0$ at a fixed $\eta$. The bottom three panels show ${\cal M}^{*}_{\%}$ as a function of $\eta$ and $e_0$ at a fixed $m$. While Fig.~\ref{fig:sanity-check2} established that the mismatch between \texttt{TaylorF2Ecck} and the other approximants increases with $e_0$ for a fixed $m$ and $\eta$, this extended analysis reveals how the mismatch depends on $m$ and $\eta$ across the parameter space.
When comparing \texttt{TaylorF2Ecck} with \texttt{TaylorF2}, ${\cal M}^{*}_{\%}$ decreases with increasing $m$ at fixed $e_0$ in the $e_0$--$m$ plane. A similar trend is observed in the $e_0$--$\eta$ plane. 
When comparing \texttt{TaylorF2Ecck} with \texttt{TaylorF2Ecc} or \texttt{TaylorF2Ecch}, ${\cal M}^{*}_{\%}$ instead increases with $m$ at fixed $e_0$ in the $e_0$--$m$ plane. A similar but relatively weaker trend is observed for $\eta$. Overall, the mismatch between \texttt{TaylorF2Ecck} and the quasicircular baseline is significantly more sensitive to changes in the mass parameters than the mismatch against the other eccentric models.}\label{fig:contour_mismatch_all}
\end{figure*}

This appendix offers an extended analysis of the mismatch percentage (${\cal M}^{*}_{\%}$), complementing the comparisons shown in Fig.~\ref{fig:contour_mismatch} of the main text. Fig.~\ref{fig:contour_mismatch_all} illustrates how ${\cal M}^{*}_{\%}$ varies with $m$, $e_0$ (evaluated at $20$~Hz), and $\eta$ when comparing our \texttt{TaylorF2Ecck} approximant against \texttt{TaylorF2Ecc} (included here for completeness), \texttt{TaylorF2Ecch}, and the quasicircular \texttt{TaylorF2} model. While Fig.~\ref{fig:sanity-check2} established that the mismatch between \texttt{TaylorF2Ecck} and the other approximants increases with $e_0$ for fixed mass parameters, this extended analysis reveals how the mismatch depends on $m$ and $\eta$ across the broader parameter space.

When comparing \texttt{TaylorF2Ecck} with \texttt{TaylorF2}, ${\cal M}^{*}_{\%}$ decreases with increasing $m$ at fixed $e_0$ in the $e_0$--$m$ plane. A similar trend is observed in the $e_0$--$\eta$ plane. This deviation from \texttt{TaylorF2} is mainly driven by the shift in the Fourier phase caused by $e_0$ corrections, which accumulates more substantially over the longer inspirals associated with lower values of $m$~\cite{Favata2022}. Consequently, the mismatch between \texttt{TaylorF2Ecck} and \texttt{TaylorF2} is significantly more sensitive to changes in the mass parameters than the mismatch between \texttt{TaylorF2Ecck} and the other eccentric models. Similar behavior is expected when \texttt{TaylorF2Ecch} or \texttt{TaylorF2Ecc} is compared against \texttt{TaylorF2}, because these eccentric approximants have similar leading-order eccentric corrections in the dominant-mode Fourier phase.

Conversely, when comparing \texttt{TaylorF2Ecck} with \texttt{TaylorF2Ecc} or \texttt{TaylorF2Ecch}, ${\cal M}^{*}_{\%}$ increases with $m$ at fixed $e_0$ in the $e_0$--$m$ plane. A similar but relatively weaker trend is observed for $\eta$. For very low eccentricities ($e_0 \lesssim 0.02$), the mismatch is relatively insensitive to changes in $m$ and $\eta$. However, at larger values of $e_0$, the dependence on these mass parameters becomes much more pronounced, especially for $m$. The trends in ${\cal M}^{*}_{\%}$ when comparing \texttt{TaylorF2Ecck} with \texttt{TaylorF2Ecch} are broadly similar to those observed in the \texttt{TaylorF2Ecck} versus \texttt{TaylorF2Ecc} comparison. Such behavior is expected because periastron advance in \texttt{TaylorF2Ecck} introduces extra harmonics and additional phase shifts in the subdominant harmonics relative to models that omit these effects.

\begin{figure*}[ht!]
    \centering
    \includegraphics{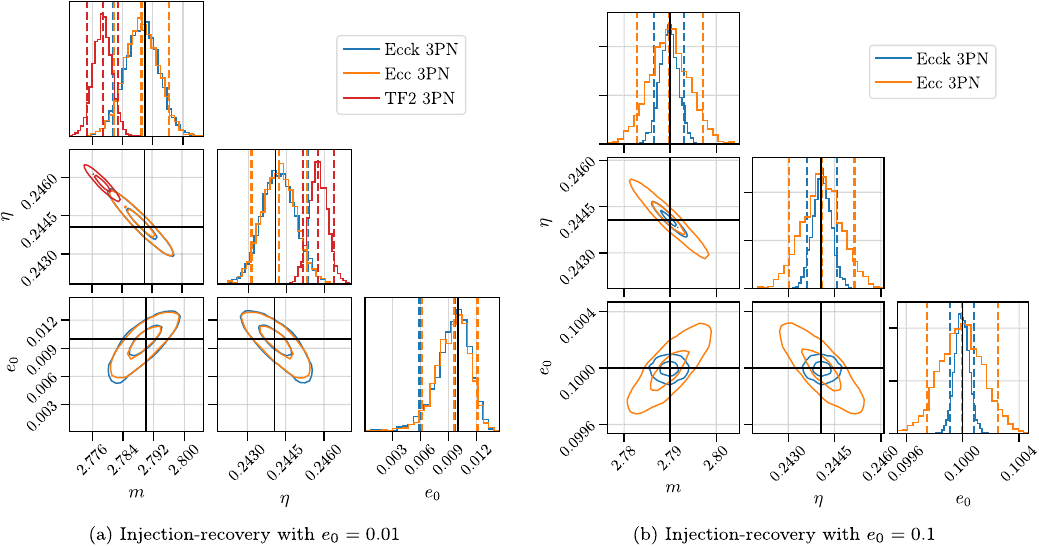}
    \caption{Corner plots showing the posterior distributions of $m$, $\eta$, and $e_0$ from injection-recovery tests using GW170817-like parameters at $20$~Hz. The panels compare the recovery of signals injected with \texttt{TaylorF2Ecck} at initial eccentricities of $e_0 = 0.01$ (panel (a)) and $e_0 = 0.1$ (panel b). Vertical solid black lines indicate the injected values. Vertical dotted lines show the median and 90\% credible intervals for three recovery models: \texttt{TaylorF2Ecck} (Ecck, blue), \texttt{TaylorF2Ecc} (Ecc, orange), and \texttt{TaylorF2} (TF2, red). The 2D contours represent the 39.3\% and 86.4\% credible regions. For these injections, Ecck and Ecc successfully recover the injected parameters within the 90\% credible intervals in both scenarios, while TF2 fails at $e_0=0.01$. At $e_0=0.1$, Ecck produces visibly narrower posteriors than Ecc.}\label{fig:injection_recovery_appendix_GW170817}
\refstepcounter{subfigure}\label{fig:injection_recovery1}
\refstepcounter{subfigure}\label{fig:injection_recovery2}
\end{figure*}

\section{Parameter estimation results from injection-recovery tests}\label{sec:appendix_injection_recovery}

This appendix presents supplementary results from our injection-recovery tests, complementing the analysis discussed in Sec.~\ref{sec:sanity_check}.

\subsection{GW170817-like injections}\label{sec:appendix_injection_recovery_GW170817}

Figure~\ref{fig:injection_recovery_appendix_GW170817} evaluates the parameter-recovery performance of \texttt{TaylorF2Ecck}, \texttt{TaylorF2Ecc}, and the quasicircular \texttt{TaylorF2} models using mock signals injected into zero noise (employing an O4 design sensitivity PSD, which yields a network SNR of $\approx 64$). For an injected eccentricity of $e_0=0.01$, the quasicircular \texttt{TaylorF2} model fails to recover the injected mass parameters within the 90\% credible interval, whereas both eccentric models succeed. As eccentricity increases to $e_0=0.1$, the inclusion of periastron advance in \texttt{TaylorF2Ecck} yields significantly tighter parameter constraints than \texttt{TaylorF2Ecc} and alters the correlation structure between the mass parameters and eccentricity. It is important to note that these observations depend heavily on this specific setup and may vary under different configurations and SNR levels.

\subsection{GW190425-like injections}\label{sec:appendix_injection_recovery_GW190425}

\begin{figure}[t!]
\centering
\includegraphics{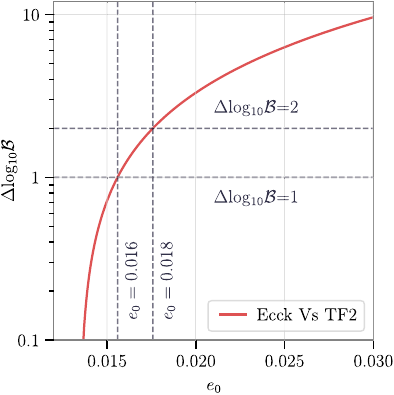}
\caption{Injection-recovery results for a simulated GW190425-like event, complementing the GW170817-like injection-recovery results shown in Fig.~\ref{fig:sanity-check2} (right panel). The comparison evaluates the eccentric \texttt{TaylorF2Ecck} model against the quasicircular \texttt{TaylorF2} model. The plot shows the relative Bayes factor $\Delta \log_{10}{\cal B}$ as a function of the injected initial eccentricity $e_0$. Horizontal dashed lines mark the standard interpretative thresholds for strong ($\Delta \log_{10}{\cal B}=1$) and decisive ($\Delta \log_{10}{\cal B}=2$) evidence of waveform distinguishability. Vertical lines indicate the injected eccentricities at which the eccentric model crosses these thresholds, $e_0=0.016$ and $e_0=0.018$, respectively.}\label{fig:injection_recovery_appendix_GW190425}
\end{figure}

To complement our GW170817-like analysis and investigate the effects of lower SNR and different mass parameters in injection recovery tests, Fig.~\ref{fig:injection_recovery_appendix_GW190425} explores waveform distinguishability for a simulated GW190425-like event (using the O4 design sensitivity PSD, yielding a network SNR of $\approx 13$). By comparing recoveries from the eccentric \texttt{TaylorF2Ecck} model against the quasicircular \texttt{TaylorF2} model, we find that the inclusion of eccentricity becomes statistically significant, reaching the conventional thresholds for strong ($\Delta \log_{10}{\cal B} \geq 1$) and decisive ($\Delta \log_{10}{\cal B} \geq 2$) evidence, at $e_0 = 0.016$ and $e_0 = 0.018$, respectively. These thresholds are higher than those obtained for the GW170817-like injection, which reached strong and decisive evidence at $e_0=0.010$ and $e_0=0.011$, respectively. This trend is qualitatively consistent with the SNR scaling of the indistinguishability criterion, where lower-SNR signals require a larger waveform mismatch, and therefore a larger initial eccentricity, to become distinguishable. The different intrinsic parameters and signal duration may also contribute to the shift in the eccentricity thresholds. As with the GW170817-like analysis, we emphasize that these specific distinguishability thresholds are highly dependent on the chosen setup and SNR levels.

\section{Parameter estimation results for inspiral-dominated events}

We present extended posterior distributions for the BNS events GW170817 and GW190425 using multiple waveform models, complementing the descriptions and plots in the main text.

\subsection{Parameter estimation results for GW170817}\label{sec:full_gw170817}

\begin{figure*}[tbp]
\centering
\includegraphics{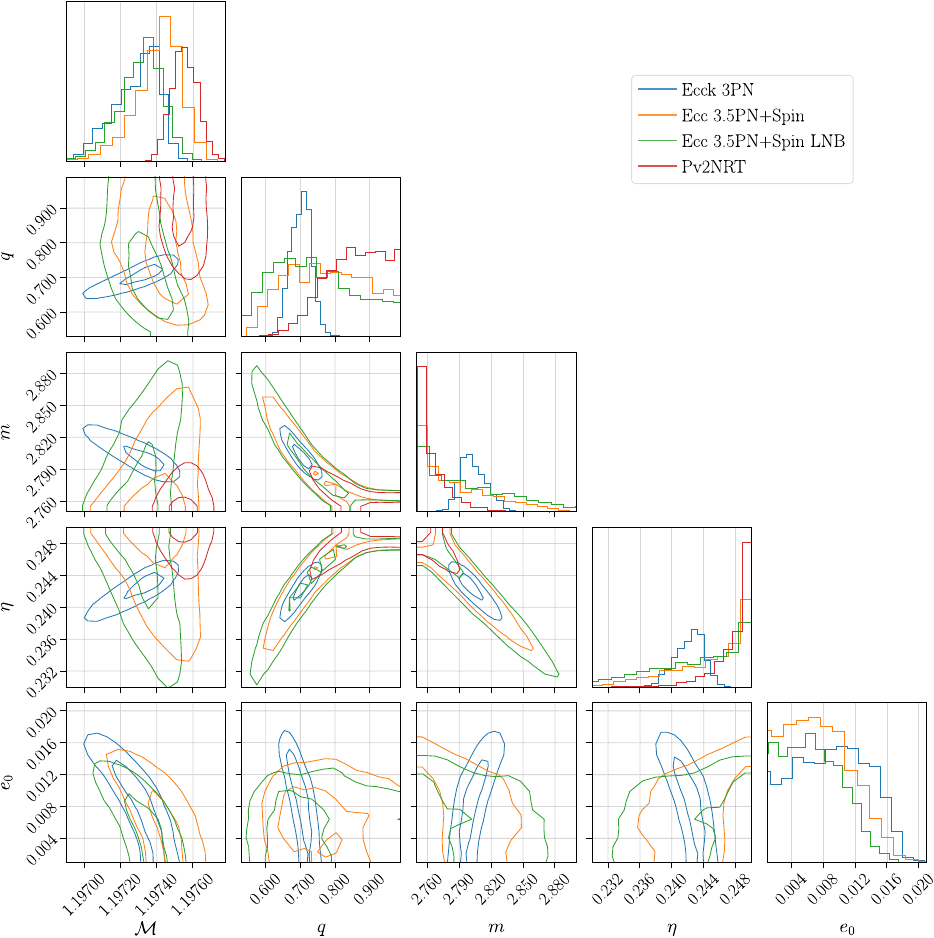}
\caption{Corner plots showing the posterior distributions for key mass parameters and eccentricity (${\cal M}$, $q$, $m$, $\eta$, $e_0$) for GW170817 under a uniform $e_0$ prior. The 2D contours represent the 39.3\% and 86.4\% credible regions. The plot compares four sets of results: our \texttt{TaylorF2Ecck} approximant (Ecck 3PN, blue), our run of the full \texttt{TaylorF2Ecc} approximant (Ecc 3.5PN+Spin, orange), the results from Lenon \textit{et al.}~\cite{LNB, LNBrepo} (Ecc 3.5PN+Spin LNB, green), and the quasicircular \texttt{IMRPhenomPv2NRT} results from the LVK GWTC-1 data release~\cite{LIGO-P1800370} (Pv2NRT, red). \texttt{TaylorF2Ecck} yields visually narrower constraints on the mass ratio $q$ compared to the 3.5PN spin-inclusive models.}\label{fig:pe_gw170817_compare}
\end{figure*}

Figure~\ref{fig:pe_gw170817_compare} contextualizes the parameter-estimation results of \texttt{TaylorF2Ecck} relative to the full \texttt{TaylorF2Ecc} model (which includes leading-order spin and 3.5PN circular contributions to the Fourier phase). We include two sets of full \texttt{TaylorF2Ecc} results: our own parameter-estimation runs and those from prior work by Lenon \textit{et al.}~\cite{LNB} (utilizing their data release~\cite{LNBrepo}). For broader context and comparison, we also include parameter-estimation results from the LVK GWTC-1 data release~\cite{LIGO-P1800370}, obtained with the quasicircular \texttt{IMRPhenomPv2NRT} model. This specific approximant is a frequency-domain phenomenological waveform that incorporates precessing spins and tidal effects calibrated to numerical relativity. While \texttt{TaylorF2Ecck} appears to provide tighter constraints on $q$ than the 3.5PN spin-corrected models, this must be interpreted cautiously because \texttt{TaylorF2Ecck} currently lacks spin effects and higher-order PN corrections to the Fourier phase. Furthermore, the slight differences observed between our \texttt{TaylorF2Ecc} posteriors and those of LNB are likely driven by different parameter-estimation settings, including their use of Markov chain Monte Carlo sampling with $f_{\rm min}=20$~Hz and an inspiral duration of $190$~s, compared to our nested-sampling setup. See Sec.~\ref{sec:analysis_techniques} for details.

\begin{figure*}[tbp]
    \centering
    \includegraphics{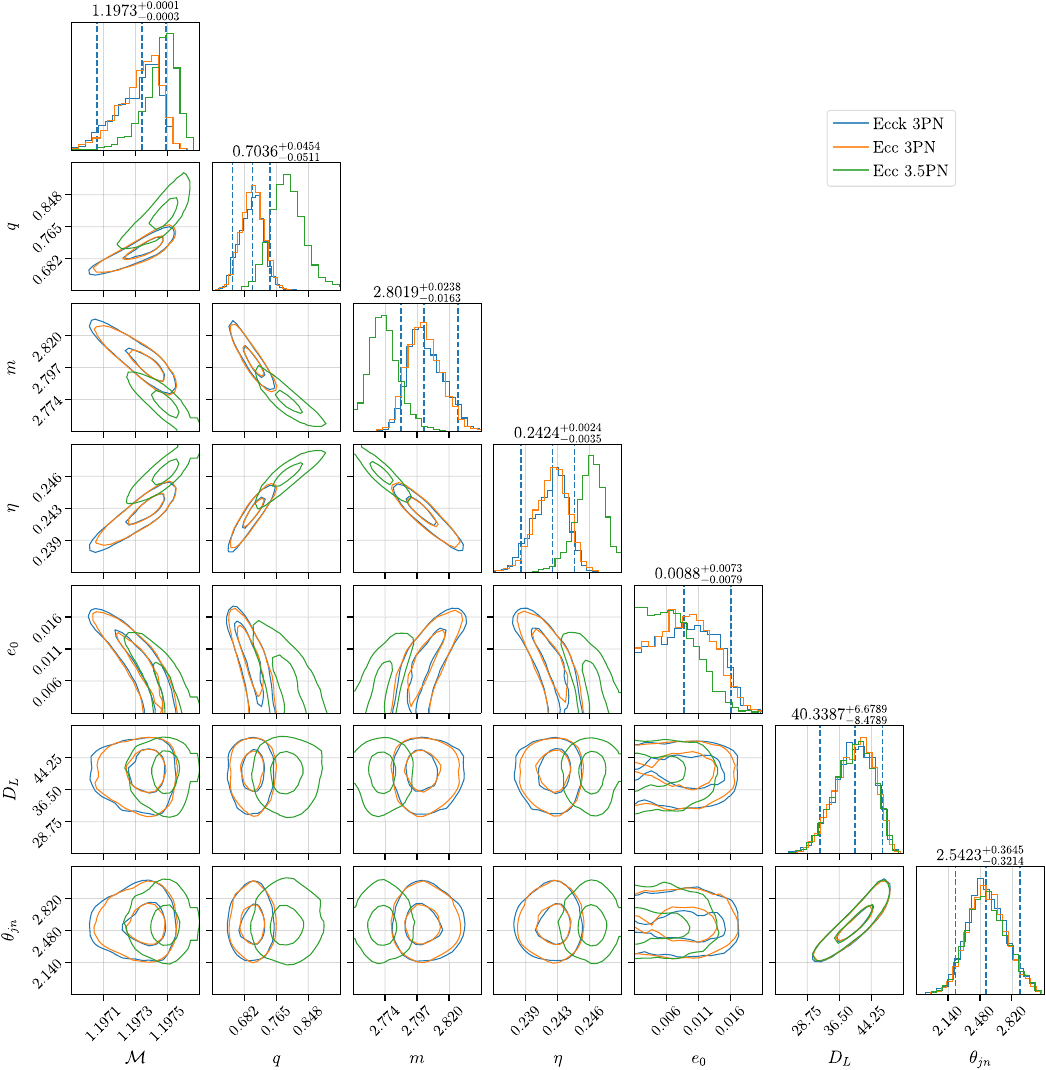}
    \caption{Extended corner plots of the posterior results for GW170817 using a uniform $e_0$ prior, complementing Fig.~\ref{fig:pe_gw170817}. Vertical dashed lines show the median and 90\% credible intervals for the \texttt{TaylorF2Ecck} posteriors, and 2D contours enclose the 39.3\% and 86.4\% credible regions. The plot compares three waveform models across intrinsic and extrinsic parameters, ${\cal M}$, $q$, $m$, $\eta$, $e_0$, $d_L$, and $\theta_{jn}$: \texttt{TaylorF2Ecck} (Ecck 3PN, blue), \texttt{TaylorF2Ecc} (Ecc 3PN, orange), and \texttt{TaylorF2Ecc} with additional 3.5PN circular Fourier phase corrections (Ecc 3.5PN, green). The posteriors for Ecck 3PN and Ecc 3PN overlap almost perfectly, while Ecc 3.5PN displays a distinct shift in the mass ratio $q$. Extrinsic parameters show no visible dependence on the choice of waveform model.}\label{fig:GW170817_full_pe}
\end{figure*}

In Fig.~\ref{fig:GW170817_full_pe}, we present extended parameter-estimation results for GW170817 with a uniform $e_0$ prior. The figure shows the posterior distributions of ${\cal M}$, $q$, $m$, $\eta$, $e_0$, $d_L$, and $\theta_{jn}$, along with their degeneracies and correlations. The plot compares \texttt{TaylorF2Ecck}, \texttt{TaylorF2Ecc} at 3PN, and \texttt{TaylorF2Ecc} with additional 3.5PN circular Fourier phase corrections, denoted by Ecck 3PN, Ecc 3PN, and Ecc 3.5PN, respectively. Because the inferred eccentricity for this event is negligible, with $e_0 \lesssim 0.016$ and the posteriors railing toward zero, and because $|\Delta \log_{10}{\cal B}|<1$, the Ecck 3PN and Ecc 3PN models produce nearly identical joint distributions. This confirms that the impact of periastron advance and higher harmonics is negligible in this low-eccentricity regime. However, as discussed in the main text, comparing Ecck 3PN directly with Ecc 3.5PN reveals a noticeable model-dependent shift in the $q$ posterior. Finally, extrinsic parameters such as $d_L$ and $\theta_{jn}$ remain consistent across all three models and show no substantial correlation with $e_0$.

\subsection{Parameter estimation results for GW190425}\label{sec:full_GW190425}

\begin{figure*}[tbp]
    \centering
    \includegraphics{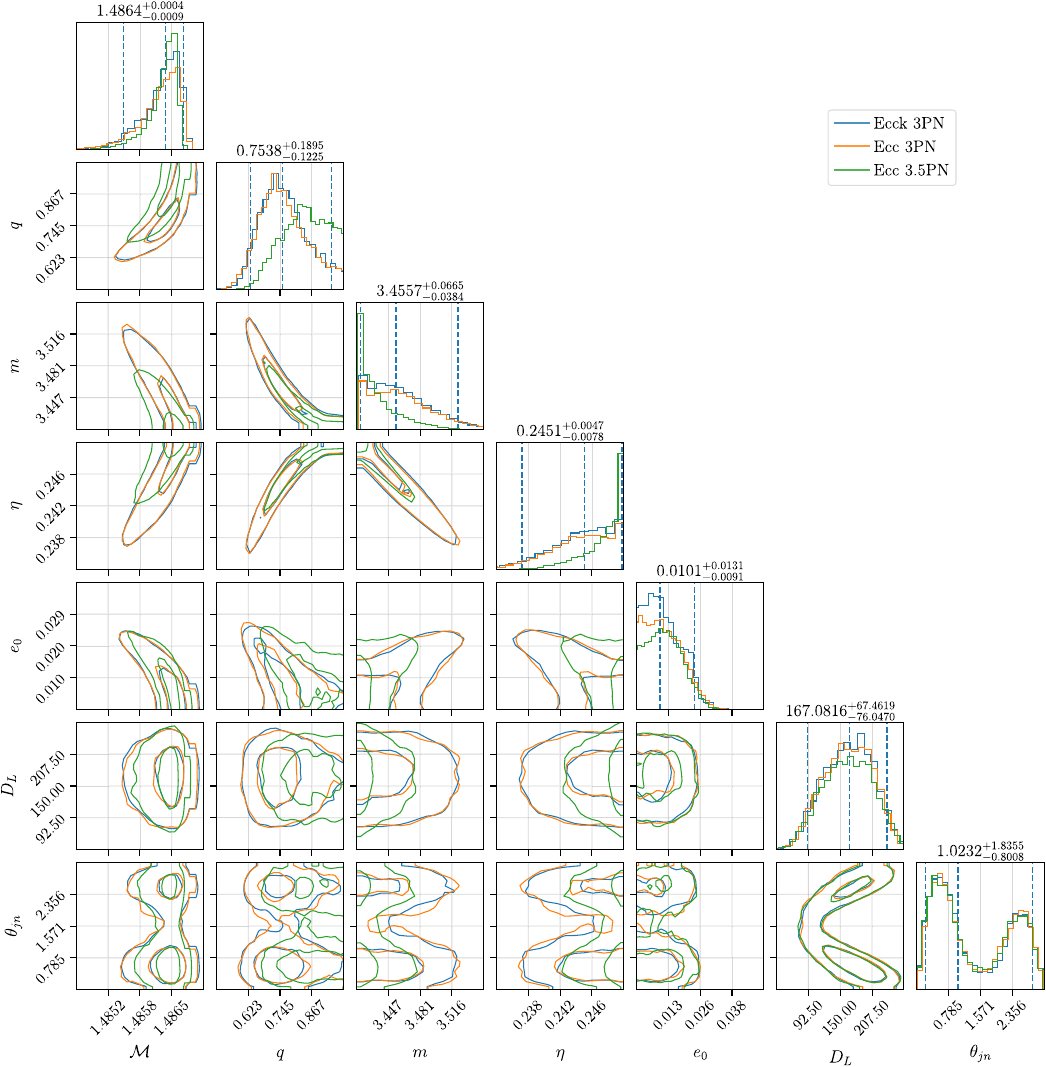}
    \caption{Extended corner plots of the posterior results for GW190425 using $f_{\rm min}=20$\,Hz and a uniform $e_0$ prior. Vertical dashed lines indicate the median and 90\% credible intervals for the \texttt{TaylorF2Ecck} posteriors, and 2D contours enclose the 39.3\% and 86.4\% credible regions. The plot compares three waveform models across intrinsic and extrinsic parameters, ${\cal M}$, $q$, $m$, $\eta$, $e_0$, $d_L$, and $\theta_{jn}$: \texttt{TaylorF2Ecck} (Ecck 3PN, blue), \texttt{TaylorF2Ecc} (Ecc 3PN, orange), and \texttt{TaylorF2Ecc} with additional 3.5PN circular Fourier phase corrections (Ecc 3.5PN, green). As with GW170817, the posteriors for Ecck 3PN and Ecc 3PN overlap closely, while Ecc 3.5PN displays a shift in the mass ratio $q$. Extrinsic parameters show no visible dependence on the choice of waveform model.}\label{fig:GW190425_full_pe}
\end{figure*}

Figure~\ref{fig:GW190425_full_pe} shows extended parameter-estimation results for GW190425 with a uniform $e_0$ prior, using the same waveform models and parameter set as in the GW170817 analysis. This analysis supports the conclusions drawn from GW170817, but with broader posteriors due to the lower SNR of GW190425 and the absence of an electromagnetic counterpart. Consequently, the constraints on both the mass ratio $q$ and eccentricity $e_0$ are weaker. Nevertheless, the inferred eccentricity remains consistent with a negligible value, with $e_0 \lesssim 0.023$, posteriors railing toward zero, and $|\Delta \log_{10}{\cal B}|<1$ when the eccentric \texttt{TaylorF2Ecck} model is compared with the quasicircular \texttt{TaylorF2} model.

As seen for GW170817, the inclusion of 3.5PN circular Fourier phase terms induces a shift in the $q$ posterior, although this shift is less pronounced for GW190425. We observe no noticeable differences in the joint posteriors involving mass and eccentricity when comparing Ecck 3PN and Ecc 3PN, confirming that periastron advance and higher harmonics have negligible impact in this low-eccentricity regime. Finally, the posteriors for luminosity distance $d_L$ and inclination $\theta_{jn}$ are consistent across all waveform models and show no substantial correlation with $e_0$.

\section{Comparison of \NoCaseChange{\texttt{pyEFPE}}, \NoCaseChange{\texttt{TaylorF2Ecck}}, and \NoCaseChange{\texttt{TaylorF2Ecc}} waveform approximants}\label{sec:comparison_pyefpe_taylorf2_family}

Figure~\ref{fig:mismatch_pyEFPE_fmax} presents a comparative analysis of the \texttt{pyEFPE} waveform model against members of the \texttt{TaylorF2} family of eccentric waveforms. Specifically, the mismatch percentage, ${\cal M}^{*}_{\%}$, is shown as a function of initial eccentricity $e_0$ for comparisons between \texttt{pyEFPE} and \texttt{TaylorF2Ecck} (at 3PN accuracy), as well as \texttt{TaylorF2Ecc} (at both 3PN and 3.5PN accuracy). The settings for this comparison are consistent with those used in the main text for the sanity checks, employing a GW170817-like binary. We impose a frequency cutoff of $f_{\rm max} = 512$~Hz, and varying this value does not alter the overall conclusions. This differs from the mismatch tests in Sec.~\ref{sec:sanity_check}, where we used $f_{\rm max}=f_{\rm LSO}$.

A key observation from Fig.~\ref{fig:mismatch_pyEFPE_fmax} is that the mismatch ${\cal M}^{*}_{\%}$ between \texttt{pyEFPE} and the \texttt{TaylorF2}-family eccentric waveforms does not approach zero even as $e_0 \to 0$. This indicates fundamental differences between the models even in the quasicircular limit. When comparing \texttt{pyEFPE} with \texttt{TaylorF2Ecc}, the mismatch is higher for the 3.5PN-accurate version than for the 3PN version, suggesting that including higher-order PN terms in \texttt{TaylorF2Ecc} increases its divergence from \texttt{pyEFPE}. In the comparison with \texttt{TaylorF2Ecck}, ${\cal M}^{*}_{\%}$ increases monotonically with $e_0$, crossing the $3\%$ threshold at $e_0 \gtrsim 0.08$. This implies a significant mismatch between \texttt{pyEFPE} and \texttt{TaylorF2Ecck} at moderate eccentricity, pointing to substantial structural differences likely driven by the partly non-analytic treatment of eccentricity and the differing harmonic content in \texttt{pyEFPE}.

\clearpage
\onecolumngrid
\noindent
\begin{minipage}{\textwidth}
\centering
\includegraphics{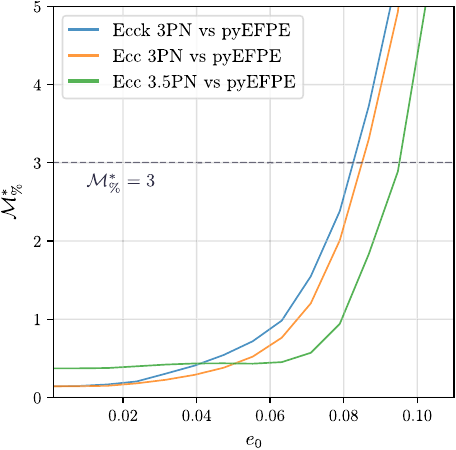}
    \captionof{figure}{Mismatch comparison between \texttt{pyEFPE} and three \texttt{TaylorF2}-family eccentric waveform approximants \texttt{TaylorF2Ecck} (Ecck 3PN), \texttt{TaylorF2Ecc} (Ecc 3PN), and \texttt{TaylorF2Ecc} (Ecc 3.5PN). The mismatch ${\cal M}^{*}_{\%}$ is plotted as a function of $e_0$ for a GW170817-like (nonspinning) binary using inspiral-only waveforms with frequency cutoffs $f_{\rm max} = 512$~Hz. In all cases, \texttt{pyEFPE} exhibits a non-negligible mismatch with the \texttt{TaylorF2}-family eccentric approximants, even when $e_0 \to 0$. Notably, the mismatch with \texttt{TaylorF2Ecck} exceeds $3\%$ for $e_0 \gtrsim 0.08$.}\label{fig:mismatch_pyEFPE_fmax}
\end{minipage}
\vspace{\textfloatsep}
\twocolumngrid

\bibliographystyle{prd-arxiv}
\bibliography{bibliography}

\end{document}